\documentclass{aa}

\usepackage{graphicx}
\usepackage{txfonts}
\usepackage{subcaption}
\usepackage{textcomp,gensymb}
\usepackage{CJKutf8}
\usepackage[colorlinks=true,linkcolor=blue,citecolor=blue,filecolor=blue,urlcolor=blue]{hyperref} % https://tex.stackexchange.com/questions/690753/aa-cls-fails-with-latex-hooks-error-under-tex-live-2023
\usepackage{color}
\usepackage{amsmath}

\definecolor{mygray}{gray}{0.6}
\definecolor{orange}{rgb}{1.0, 0.4, 0.0}
\definecolor{purple}{rgb}{0.75,0.25,0.75}

\newcommand{\hr}{\mbox{HR 8799}\xspace}
\newcommand{\pds}{\mbox{PDS 70}\xspace}
\newcommand{\MSun}{\mbox{M$_\odot$}\xspace}

\def\showdiff{0} % 0 for clean version, 1 for version with changes
\newcommand{\diffrem}[1]{\if\showdiff1\textbf{\sout{#1}}\fi}
\newcommand{\diffadd}[1]{\if\showdiff1\textbf{#1}\else#1\fi}
\NewDocumentEnvironment{diffenable}{+b}{\if\showdiff1#1\fi}

\title{Stability of a cluster-disrupted mean-motion resonance (chain) in \hr and \pds}

\author{B. Maas\inst{1} \and S. Huang (\begin{CJK*}{UTF8}{gbsn}黄硕\end{CJK*})\inst{1,2} \and S. Portegies Zwart\inst{1}}

\institute{Leiden Observatory, Leiden University, P.O. Box 9513, 2300 RA Leiden, The Netherlands 
\and
Department of Astronomy, Tsinghua University, 100084 Beijing, China
}
\date{Received N/A / Accepted N/A}

\abstract{\hr is a planetary system in which the four observed planets potentially form a mean-motion resonance chain. Although potentially forming a resonance chain, it is not clear from the observations if they are in mean-motion resonance. Similarly, \pds is a planetary system in which the two observed planets are potentially in mean-motion resonance.} %Context
{We study the stability of \hr and \pds under external perturbations to test how it responds under resonance and under mean-motion resonance.} %Aims
{We integrate the equations of motion of the planets in \hr and \pds starting with a system either in resonance or in mean-motion resonance and study the stability of \hr and \pds in isolation and in a star cluster. In the star cluster, we take the effects of passing stars into account. The dynamics of the star cluster are resolved using the Lonely Planets module in AMUSE.} %Methods
{\hr and \pds in mean-motion resonance are stable, whereas in non-resonance they dissolve on timescales of $0.303\pm0.042$ Myr and $1.26\pm0.25$ Myr, respectively. In a cluster, the non-resonant planetary system of \hr is slightly more stable than in isolation, but still dissolves on a timescale of $0.300\pm0.043$ Myr, whereas the resonant planetary system remains stable for at least $0.71$ Myr. In contrast, the non-resonant planetary system of \pds is approximately equally stable in a cluster compared to isolation and dissolves on a timescale of $1.03\pm0.20$ Myr, whereas the resonant \pds planetary system remains stable for at least $0.83$ Myr.} %Results
{Considering the more stable solutions of mean-motion resonance for \hr, we argue that the planetary system was born in mean-motion resonance and that the mean-motion resonance stayed preserved. If \hr was not born in resonance, the probability that it survived until the present day is negligible. Similarly, we argue that the planetary system of \pds was probably born in mean-motion resonance and that the mean-motion resonance stayed preserved. We also find that it is almost possible for planetary systems with a broken mean-motion resonance chain to survive longer in a perturbing cluster environment compared to isolation.} %Conclusions


\begin{document}

\maketitle

\nolinenumbers

\section{Introduction}
%Define MMR
Two planets around a star are in orbital resonance if the ratio of their orbital periods is near a ratio of integers. A special case of orbital resonance is mean-motion resonance (MMR), where the point of the closest approach between the planets librates around a fixed angle \citep{Laplace1799}. Multiple connected pairs of planets in MMR form an MMR chain. Well-known examples of MMR chains are the TRAPPIST-1 system \citep{Gillon2016,Gillon2017,Luger2017}, the Jovian moons Io, Europa and Ganymede \citep{Laplace1824}, Kepler-80 \citep{MacDonald2016}, and HD 110067 \citep{Luque2023}.

%MMR stability concerns
Planetary migration naturally leads to stable MMR chains \citep[and reference therein]{Huang2021,TeyssandierEtal2022}, though approximately 80\% of all systems do not retain or end up in an MMR chain \citep{HuangOrmel2022i}. This includes the Solar system\footnote{Solar System Dynamics. 2024-10-23. Planetary Physical Parameters. \url{https://ssd.jpl.nasa.gov}}. The TRAPPIST-1 system and the Jovian moons Io, Europa and Ganymede have an age in the order of several billions of years \citep{Burgasser2017,Malhotra1991}. These ages suggest that MMR chains may play a role in the long-term stability of planetary systems, and oppose models where MMR chains form during planet formation and destroyed soon after formation, e.g. the formation-then-break models of \citet{IzidoroEtal2022,GriveaudEtal2024,LiEtal2024,LiveoakMillholland2024}. An MMR is lost through post-formation perturbations such as collisions or flybys. The timescale at which collisions occur increases for wider orbits. There are case studies of the effects of stellar flybys on MMR chains \citep[e.g.][]{Zink2020}. A more general investigation by \citet{Charalambous2025} shows that these effects depend on, among others, the architecture of the mean-motion resonance chain and the orientation of the passing star with respect to the planetary system. Furthermore, this work shows a dichotomy between instantaneous and delayed disruptions to the planetary system.

\begin{figure}
    \centering
    \begin{subfigure}{0.49\columnwidth}
        \includegraphics[keepaspectratio=true,width=\textwidth]{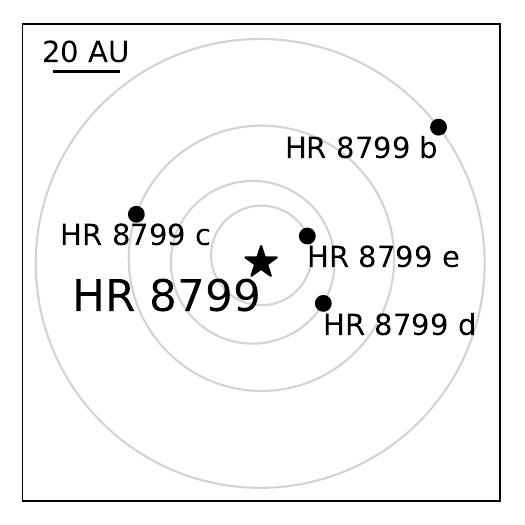}
        \caption{\hr}
        \label{fig:doodle_hr}
    \end{subfigure}
    \hfill
    \begin{subfigure}{0.49\columnwidth}
        \includegraphics[keepaspectratio=true,width=\textwidth]{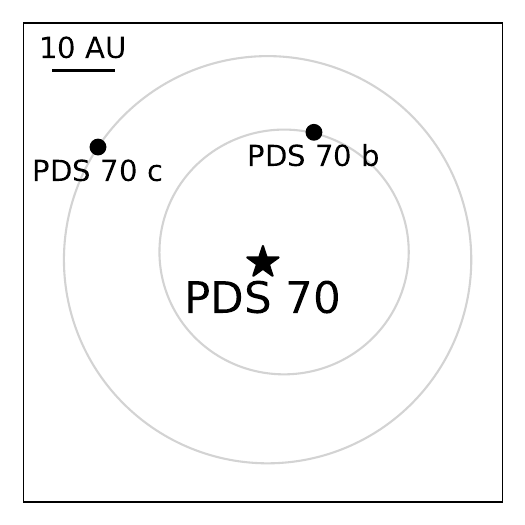}
        \caption{\pds}
        \label{fig:doodle_pds}
    \end{subfigure}
    \caption{Overviews of the planetary systems of \hr and \pds. These overviews use the resonant initial conditions created in section \ref{methods_initial_conditions}.}
\end{figure}

%Introduce HR and PDS
We investigate the planetary systems of \hr and \pds, and the potential MMRs therein. The \hr system is composed of a $1.43_{-0.07}^{+0.06}$ \MSun star orbited by four planets \citep{Marois2008,Marois2010,Sepulveda2022} (see figure \ref{fig:doodle_hr}). It has an age between $20$ Myr and $50$ Myr. The system has previously been observed using direct imaging. However, none of the planets have completed a full orbit throughout the range of direct images, leading to some orbital parameters of the planets being poorly constrained \citep{GozdziewskiMigaszewski2020}. \pds contains two observed planets that orbit a $0.76\pm0.02$ \MSun star \citep{Keppler2018,Muller2018,Haffert2019} with semi-major axes in the same order of magnitude as the planets of \hr \citep{Wang2021} (see figure \ref{fig:doodle_pds}). It has an age of $5.4\pm1.0$ Myr. Each pair of adjacent planets in \hr and \pds is near a 2:1 integer period ratio, which indicates a potential 8:4:2:1 MMR chain in \hr and a potential 2:1 MMR in \pds. The potential MMR chain cannot be confirmed through observations within our lifetimes due to the long orbital periods of the planets. However, we can use numerical simulations to observe the stability of the two systems in isolation or while being perturbed by neighbouring cluster stars, and under the assumption that the systems are in MMR or not.

In this paper, we investigate the presence and stability of MMR in the planetary systems of \hr and \pds. We do so by comparing numerical simulations where MMR is or is not present in the planetary systems, and where the planetary systems are in isolation or perturbed by neighbouring cluster stars. Section \ref{methods} contains the setup for the numerical simulations. The results of these simulations are found in section \ref{results} and discussed in section \ref{discussion}. Finally, section \ref{conclusion} summarises our conclusions.

\section{Methods}
\label{methods}
%Intro methods
Simulating a stellar cluster with several planetary systems is computationally expensive. The total simulation time and the time step are determined by the long dynamical timescales of the stellar cluster and the short dynamical timescales of the planetary systems, respectively. The methodology of this work is based on Lonely Planets \citep{cai2017} to get a reasonable total simulation time.

%Outline Lonely Planets
Lonely Planets solves the problem of perturbations on planetary systems by encounters with neighbouring cluster stars by splitting the problem into two stages. First, we simulate a stellar cluster without any planetary systems and we record the interaction history of the stars. Then, we assign an \hr system or a \pds system to a cluster star that matches the mass of the host star of \hr or \pds and we integrate the planetary system along with the interaction history of its assigned host star. By splitting the problem into these two stages, we can make more accurate comparisons between different planetary systems because these planetary systems experience identical interaction histories. If the systems experienced slightly different interaction histories, the final conditions may be drastically different as these systems of three or more bodies are chaotic \citep{Miller1964}. \citet{Nesvorn2012} and \citet{Clement2019} demonstrate the quantitative and qualitative effects that chaos can have on numerical simulations involving giant planets. Furthermore, reusing the interaction history of the stellar cluster saves us computing resources and allows us to finish simulations in a shorter amount of time. All used scripts (except for generating the initial conditions), the initial conditions and the resulting data are available on \url{https://doi.org/10.5281/zenodo.15274962}. %TODO: Publish data repository before resubmission, check for anything to do
We delve more into the details of our implementation of Lonely Planets in the following subsections.

\subsection{Initial conditions}
\label{methods_initial_conditions}

%Description cluster simulation
We simulate a stellar cluster using the integrator Ph4 \citep{portegieszwart2018} in the AMUSE simulation framework \citep{PortegiesZwart2008,PortegiesZwart2013,Pelupessy2013,portegieszwart2018,amuselatest}. We initialise the stellar cluster similar to the Solar birth cluster \citep{portegieszwart2019}: 5000 stars in a Plummer sphere \citep{Plummer1911} with a Plummer radius of 0.7 pc, and with a Kroupa initial mass function between 0.08 \MSun and 100 \MSun \citep{Kroupa2002}. We place the cluster in a Milky Way potential \citep{Bovy2015} such that the centre of the cluster follows the modern-day orbit of either \hr or \pds. We create the initial conditions using the position, parallax and motion of \hr and \pds from the Gaia third data release \citep{Prusti2016,Vallenari2023} and from \citet{gontcharov2006}. These orbits are relative to the Sun's galactic orbit \citep{Reid2014,Karim2016}. We simulate for 10 Myr with a step size of 1 kyr. This period of time is a balance between the cluster's ability to perturb declining, e.g. through stellar mass loss \citep{Vink2008}, and providing the planetary systems enough time to be perturbed. We record the positions, velocities, masses and radii of the cluster stars after each step of the simulation.

%Preselection of host stars
We preselect the host stars in the cluster by selecting the 50 stars whose masses are the closest to the mass of \hr and \pds. This typically results in stars whose original mass was between 1.34 \MSun and 1.52 \MSun for \hr, and 0.74 \MSun and 0.78 \MSun for \pds. We set the masses of the preselected host stars to be equal to the mass of \hr or \pds before the simulation starts. Otherwise, we would need to change the mass of the host star after the simulation has finished or scale the planetary system to fit the host star's mass. This would lead to undesired side effects, like an incorrect interaction history or an altered interaction cross-section. We give \hr and \pds each their own stellar cluster to prevent competition between their preselections and to minimise any influence that setting the mass of the host stars may have. The effects on the initial mass function are negligible as the number of changed host star masses is small enough and the density of the initial mass function around the masses of \hr and \pds is high enough such that all changes to the host star masses are small.

%Mass loss
Mass loss due to stellar evolution will become important for high-mass stars on the timescale of the simulation \citep{Vink2008}. We account for this mass loss by changing the mass of the stars during the cluster simulation using the stellar evolution solver SeBa \citep{portegieszwart1996,Toonen2012}.

\begin{figure}
    \centering
    \begin{subfigure}{\columnwidth}
        \includegraphics[keepaspectratio,width=\linewidth]{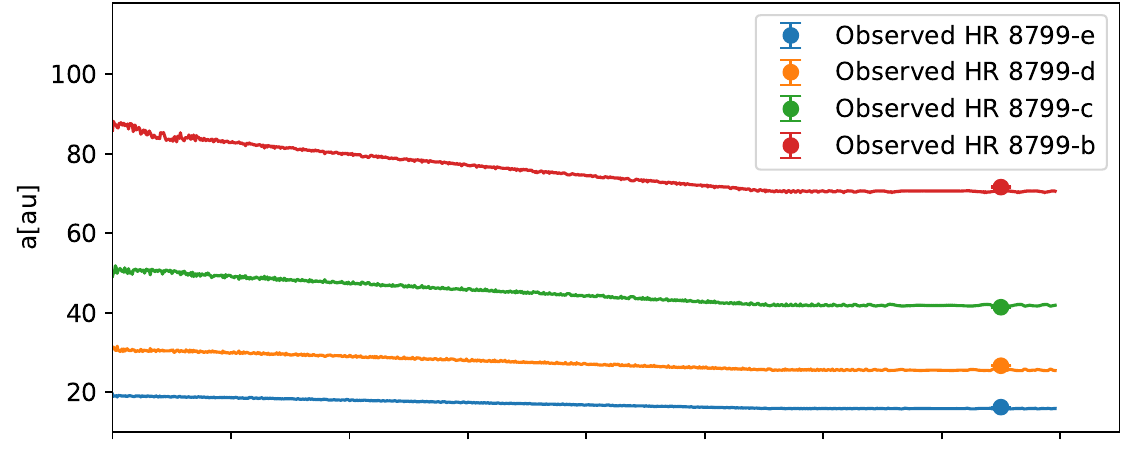}
        \caption{The semi-major axes of the four planets.}
    \end{subfigure}
    \begin{subfigure}{\columnwidth}
        \includegraphics[keepaspectratio,width=\linewidth]{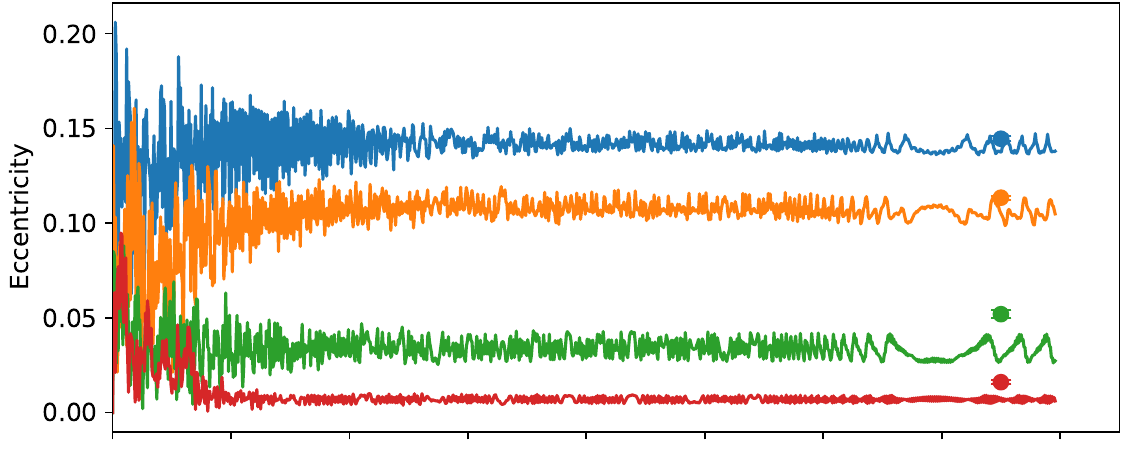}
        \caption{The eccentricities of the four planets.}
    \end{subfigure}
    \begin{subfigure}{\columnwidth}
        \includegraphics[keepaspectratio,width=\linewidth]{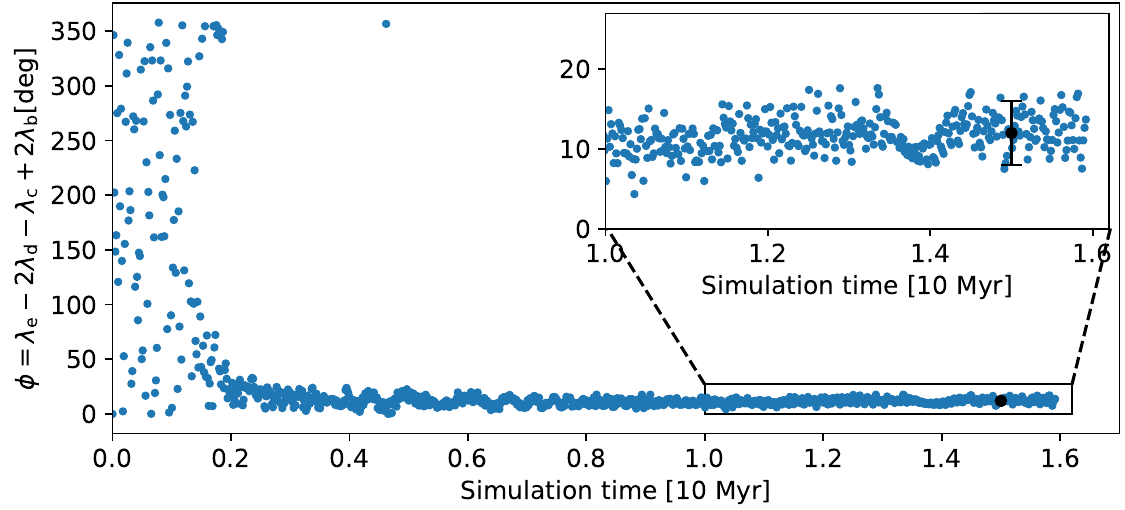}
        \caption{The 4-body resonance angle of the planetary system.}
    \end{subfigure}
    \caption{The semi-major axes, the eccentricities and the 4-body resonance angle during the initialisation of \hr's initial conditions. The data points with error bars indicate the observed values of the semi-major axes, the eccentricities and the 4-body resonance angle \citep{GozdziewskiMigaszewski2020}.}
    \label{fig:init}
\end{figure}

%Resonant initial conditions
We create initial conditions of \hr's planetary system with an MMR chain by initially placing the four planets slightly wider than the 2:1 period ratio from observations \citep{GozdziewskiMigaszewski2020}. We then push the outermost planet inwards to ensure convergent migration similar to \citet{TamayoEtal2017}, while dampening the eccentricities to match the observations. We integrate this system using REBOUND \citep{Rein2012} with WHFAST \citep{Rein2015whfast}. As shown in figure \ref{fig:init}, the system achieves MMR through the inwards migration of the outer planet, so the MMR chain is formed outside-in. We go through a similar process to create \pds's initial conditions from observations of \pds \citep{Wang2021}.

%Non-resonant initial conditions
We create non-resonant initial conditions for \hr and \pds by modifying their resonant initial conditions. We consider each planet as a binary component, whereas the other component is a virtual body. This virtual body represents the combination of the host star and the planets closer to the host star than the considered planet and takes on their combined mass, centre-of-mass position and centre-of-mass velocity. We use a virtual body instead of just the host star itself to account for the effects that close-in planets have on the host star. We then decompose the binary of the planet and the virtual body into orbital elements, randomise the mean anomaly, and recompose the modified orbital elements into a binary. Finally, we adjust the positions and velocities of the considered planet and the bodies represented by the virtual body according to the changes in position and velocity in the modified binary.

%Initial conditions for planetary systems
The resulting initial conditions can be found in appendix \ref{appendix_initial_conditions} in table \ref{table:initial_conditions} for \diffadd{both} resonant initial conditions and \diffrem{table A.2 for} non-resonant initial conditions. We can verify whether the initial conditions are in MMR or contain an MMR chain using resonance angles (see appendix \ref{resonance_angles}). We assign copies of both kinds of initial conditions of both planetary systems to each of the preselected host stars.

\subsection{Stellar encounters}
\label{methods_stellar_encounters}
%Fixed number of neighbours
We take the encounters from the simulated cluster from section \ref{methods_initial_conditions}. Determining when a star should be considered an encounter is a complex problem; every star always affects any planetary system due to the infinite range of gravity. Furthermore, the effect of an encounter depends on multiple parameters of the perturbing star, for example: the mass of the perturber, how close the perturber gets, and the relative velocity of the perturber \citep{Hut1983}. We consider a fixed number of six neighbouring stars as perturbers at any given time.

%Nearest neighbours selection
There are multiple ways to choose the neighbours. The simplest way is to choose the stars with the shortest distance to the host star. However, this method introduces the risk that a strong perturber may be obscured by a couple of closer low-mass stars. A more refined approach is to choose the stars which exert the strongest gravitational force on the host star. This mitigates the screening of high-mass stars by low-mass stars, but a stronger gravitational force does not necessarily lead to a stronger perturbation as the host star and its planets experience a near-identical force from the perturber. Therefore, we choose to select perturbing neighbours based on the difference in the force exerted on the host star compared to its planets, i.e. the gradient of the gravitational force at the host star.

%Selection quantities
We perform our selection using a $k$-nearest neighbours search \citep{Fix1951} by maximising the selection quantity $q$. For nearest neighbour and gravitational force selection, these quantities are $q_{nn}=1/r$ and $q_F=M/r^2$, respectively, where any quantities irrelevant to the selection are removed (e.g. constants and properties of the host star). Selecting on distance or gravitational force only involves two bodies (the host star and the candidate neighbour), but the gravitational force gradient-based selection involves at least three bodies as we now have to consider the planetary system of the host star as well. For simplicity, we have reduced this scenario to the host star, the candidate neighbour and the outermost planet being spatially aligned with the planet between the stars at a distance $d$ from the host star\footnote{We can also involve the inclination of the planet with respect to the neighbouring star or integrate the force difference over an entire orbit to get more accurate results. However, we prefer this simpler method}, ignoring any other planets. This leads to the selection quantity $q_{\nabla F}$:
\begin{equation}
    q_{\nabla F}=\frac{M}{(r-d)^2} - \frac{M}{r^2}.
\end{equation}
For human intuition, it can be useful to collapse the equation into a single term. This can be done by assuming that the planet is much closer to its host star than to the candidate neighbour ($d<<r$):
\begin{equation}
\label{eq:neighbour_selection_force_difference}
    q_{\nabla F}=-Md\frac{\frac{1}{(r-d)^2}-\frac{1}{r^2}}{-d}\approx-Md\frac{\partial\frac{1}{r^2}}{\partial r}=\frac{2Md}{r^3}\sim \frac{M}{r^3}
\end{equation}
We can again remove the factors that are not properties of the candidate neighbour to get $q_{\nabla F}\approx M/r^3$. The derivative in equation \ref{eq:neighbour_selection_force_difference} shows that selection based on the difference between the gravitational force exerted on the host star and the gravitational force exerted on its planets is approximately equivalent to selection based on the gradient of the gravitational force exerted on the host star.

%Interpolation
The dynamical timescale of a planetary system is much smaller than the dynamical timescale of the stars in the stellar cluster. Therefore, we have to either record the data of the stars in the cluster at an unnecessarily high rate or reconstruct the paths of the stars by interpolating the recorded data. We choose to interpolate the recorded data to save on computer resources as we deem the recorded cluster data to be recorded densely enough. We perform this interpolation by fitting a 4\textsuperscript{th} order polynomial between adjacent data points of each of the three components of both the positions and the velocities of the stars.

\subsection{Integrating the planetary systems}
\label{methods_integrating_systems}
%Symple
We use the N-body code Symple, which is part of AMUSE\footnote{Documentation for Symple will be added in a future iteration of the AMUSE textbook \citep{portegieszwart2018}}, to integrate the planetary systems because of its symplectic nature and because the version of REBOUND used before does not support multiple strong perturbers. We configured Symple to use 6\textsuperscript{th} order integration with a time step parameter of $0.01$. We run the simulations for 10 Myr to match the run time of the cluster simulation. We save a snapshot every 1 kyr to sample the orbit of the least bound outer planet every couple of orbits. These simulations consist of both isolated runs and runs perturbed by neighbouring cluster stars.

%Neighbour handling
At the start of the run, we load a planetary system and, if the runs is a perturbed run, add its neighbouring stars as determined in section \ref{methods_stellar_encounters} relative to the host star of the planetary system. Any present neighbours will deviate from their paths in the cluster as most of the cluster is now absent, so we update their positions and velocities every 100 yr using the interpolated recorded data from the cluster to correct the deviations. Whenever another star overtakes a neighbour, the neighbour is removed and this other star becomes a neighbour.

%Stopping conditions
An encounter with a neighbouring star or an internal dynamical perturbation may be severe enough that one or more planets are lost. In that case, we stop the run. We detect these events after each time step by checking if any of the planets is both energetically unbound (i.e. its kinetic energy exceeds its gravitational potential energy imposed by the planetary system) and at least 100 times the initial Hill radius of the initial outermost planet away from the host star. A planet that fulfils both of these conditions could never return to the planetary system. If any planet meets these two conditions, the run is ended and the planetary system is considered (partially) destroyed.

\section{Results}
\label{results}

\subsection{Survival time in isolation}
\label{survival_time_isolation}
%Intro survival time in isolation
For both \hr and \pds, we run a single simulation with resonant initial conditions and 50 simulations with 50 different non-resonant initial conditions. We run these simulations up to 10 Myr. The planetary systems are in isolation. The reason for only running a single simulation with isolated resonant initial conditions is that we only have one set of resonant initial conditions. Integrating copies of this set of initial conditions is not useful as isolated simulations of identical initial conditions will produce identical results. We present the results of these simulations in terms of the survival rate over time, $f_s(t)$, separately for \hr and \pds in the following subsections. These results show that the resonant initial conditions of both \hr and \pds created in section \ref{methods_initial_conditions} are stable enough to survive the entire 10 Myr in isolation.

\subsubsection{\hr}
\label{survival_time_isolation_hr}

\begin{figure*}
    \centering
    \includegraphics[keepaspectratio=true,width=\textwidth]{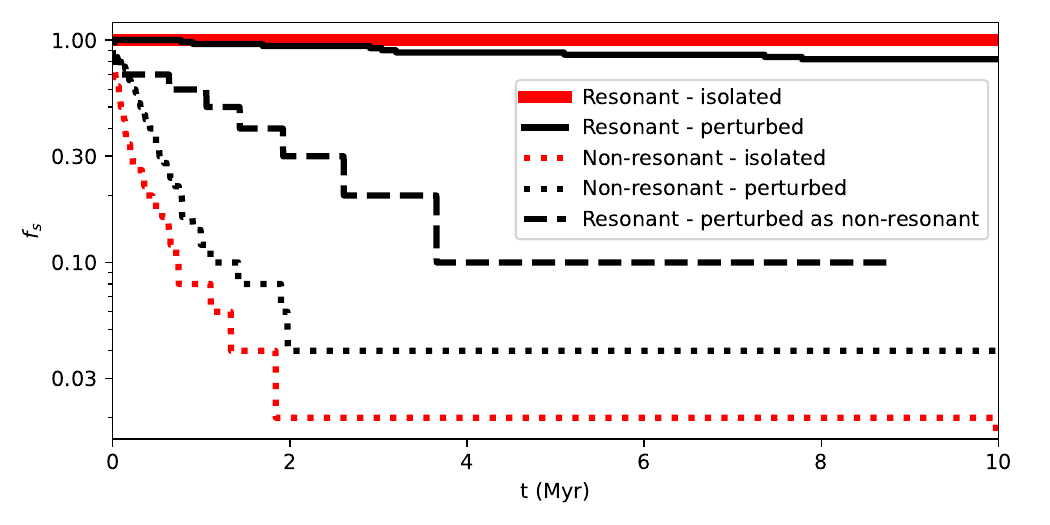}
    \caption{The fraction of surviving planetary systems over time ($f_s(t)$) for \hr. The figure shows the isolated resonant system, the 50 perturbed resonant systems, the 50 isolated non-resonant systems, and the 50 perturbed non-resonant systems. In addition, a reinterpretation of the perturbed resonant systems that lost resonance is shown. These systems are shifted in time such that the moment they lose resonance is at $t=0$. This subset constitutes 10 \hr systems. All sets of systems display distinct behaviour from one another. The perturbed resonant systems that lost resonance dissolve slower than the non-resonant systems. The curve showing reinterpreted perturbed resonant systems is cut off at 8.8 Myr as we do not have data for the last 1.2 Myr. Later analysis in section \ref{discussion} shows in table \ref{table:survival_stats} that the isolated resonant curve and the perturbed resonant curve do not have a statistically significant difference as the difference between $0$ and the value for $\lambda$ for the perturbed resonant systems is not statistically significant.}
    \label{fig:survival_rate_hr}
\end{figure*}

\begin{figure}
    \centering
    \includegraphics[keepaspectratio=true,width=\columnwidth]{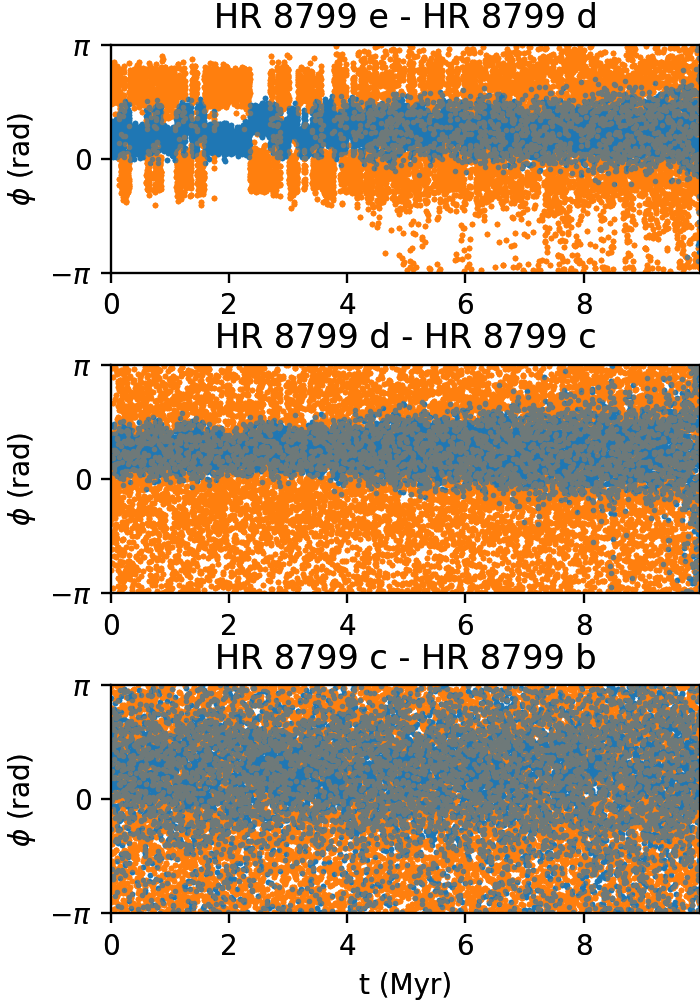}
    \caption{The 2-body resonance angles over time of an isolated non-resonant \hr system that is near MMR, starting with the inner planet pair (\hr e and \hr d) and ending with the outer planet pair (\hr c and \hr b). The colours show the use of the argument of pericentre of either the inner planet (blue) or the outer planet (orange). The plots show that the planets are near MMR as the resonance angles are librating. These arguments of pericentre are more noisy towards the outer planets as these planets have lower eccentricities. Therefore, the argument of pericentre is more sensitive to change and becomes harder to determine.}
    \label{fig:oor_near_resonance_isolated_hr_2br}
\end{figure}

\begin{figure}
    \centering
    \includegraphics[keepaspectratio=true,width=\columnwidth]{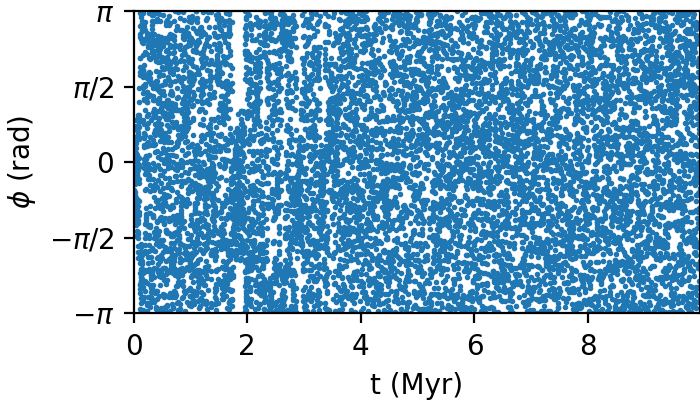}
    \caption{The 4-body resonance angle over time of an isolated non-resonant \hr system that is near MMR. The plot shows that the planets do not form an MMR chain as the resonance angle does not librate.}
    \label{fig:oor_near_resonance_isolated_hr_4br}
\end{figure}

%All survive in MMR, none survive out of MMR, but one nearly does due to being near MMR
The survival rates over time for isolated \hr systems are shown in figure \ref{fig:survival_rate_hr}. The resonant system survives until the cutoff time of 10 Myr, while none of the 50 non-resonant systems do. One non-resonant system, however, almost survives until 10 Myr, but dissolves in the last 30 kyr. When inspecting its resonance angles (see figure \ref{fig:oor_near_resonance_isolated_hr_2br} and \ref{fig:oor_near_resonance_isolated_hr_4br}), we see that its randomised mean anomalies are close to MMR. Although near MMR, the system does not form an MMR chain. The semi-major axis differences $\Delta$ in terms of mutual Hill radii $R_H$ as described by \citet{Smith2009} (who use $\beta$ instead of $\Delta$) are $\Delta_{e-d}=3.20R_{H,e-d}$, $\Delta_{d-c}=2.80R_{H,d-c}$ and $\Delta_{c-b}=3.98R_{H,c-b}$ for the consecutive planets pairs of \hr e through \hr b. These semi-major axis differences are well below the instability criterium of $\Delta<10R_H$ of \citet{Chambers1996} and $\Delta\lesssim8.4$ of \citet{Smith2009} for multiplanetary systems, and mostly below the instability criterium of $\Delta<2\sqrt{3}R_H\approx3.46R_H$ of \citet{Gladman1993} (except for $\Delta_{c-b}$) for systems with two planets. Therefore, the instability of the non-resonant systems is not unexpected.

\begin{figure*}
    \centering
    \includegraphics[keepaspectratio=true,width=\textwidth]{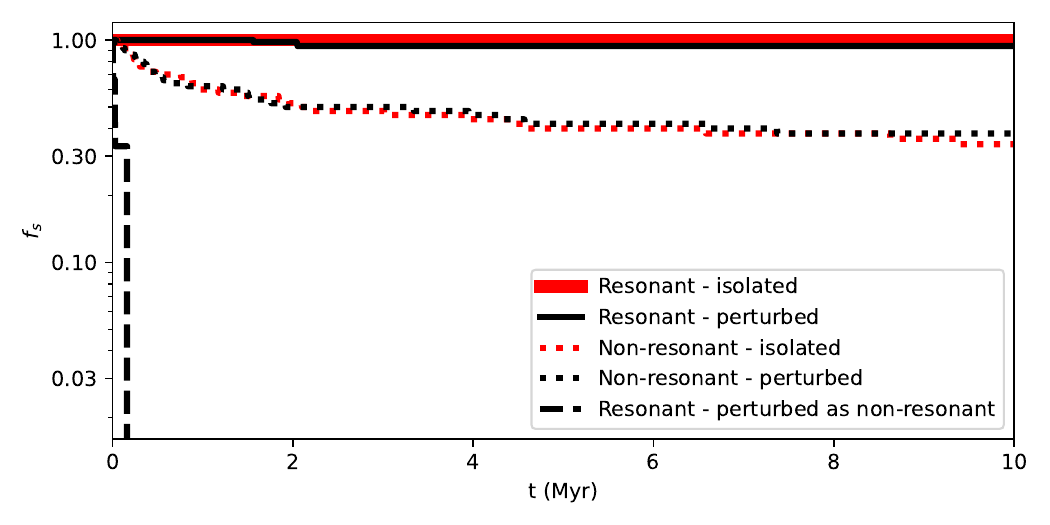}
    \caption{The fraction of surviving planetary systems over time ($f_s(t)$) for \pds. The figure shows the isolated resonant system, the 50 perturbed resonant systems, the 50 isolated non-resonant systems, and the 50 perturbed non-resonant systems. In addition, a reinterpretation of the perturbed resonant systems that lost resonance is shown. These systems are shifted in time such that the moment they lose resonance is at $t=0$. This subset constitutes 3 \pds systems. The isolated non-resonant systems and the perturbed non-resonant systems are not distinct from each other. The perturbed resonant systems that lost resonance dissolve much faster than the non-resonant systems. Later analysis in section \ref{discussion} shows in table \ref{table:survival_stats} that the isolated resonant curve and the perturbed resonant curve do not have a statistically significant difference as the difference between $0$ and the value for $\lambda$ for the perturbed resonant systems is not statistically significant.}
    \label{fig:survival_rate_pds}
\end{figure*}

\subsubsection{\pds}
%All survive in MMR, 17 out of 50 survive out of MMR, mostly because they're near MMR
The survival rates over time for isolated \pds systems are shown in figure \ref{fig:survival_rate_pds}. The single resonant system survives until the cutoff time of 10 Myr, but unlike \hr, 17 out of 50 non-resonant systems survive until the cutoff time of 10 Myr. Similar to the 2-body resonance angles of one non-resonant \hr system (figure \ref{fig:oor_near_resonance_isolated_hr_2br}), the 2-body resonance angles of 15 of these 17 non-resonant \pds systems indicate the systems are near MMR. This increased fraction of randomly generated MMRs for \pds can be explained by the number of planets in each system: \hr has four planets and \pds has two planets. Therefore, \hr requires three planets to happen to line up with their inward neighbour to be fully in MMR, while \pds only requires one planet to do so. The steady decay of non-resonant systems is not unexpected, as the semi-major axis difference $\Delta=2.97R_H$ is below the instability criterium of $\Delta<2\sqrt{3}R_H\approx3.98R_H$ of \citet{Gladman1993} for systems with two planets.

\subsection{Survival time in a cluster}
\label{survival_time_cluster}
%Intro survival time in a cluster
For both \hr and \pds, we run 50 resonant and 50 non-resonant simulations where the planetary systems are perturbed by their neighbouring stars. We compare the behaviour of these perturbed systems against the behaviour of the isolated systems from section \ref{survival_time_isolation}. In the resonant simulations, we copy the resonant initial conditions 50 times and assign each copy to a unique host star (see section \ref{methods_initial_conditions}). In the non-resonant simulations, we assign the 50 different non-resonant initial conditions from section \ref{survival_time_isolation} to the same set of unique host stars. We present the results of these perturbed simulations in terms of the survival rate over time, $f_s(t)$, separately for \hr and \pds in the following subsections. We present the subset of the perturbed resonant systems that lost their resonance to compare with the perturbed non-resonant systems. We do this by reinterpreting these systems as non-resonant systems, where time is shifted such that the moments each system loses resonance all line up with each other.

\begin{figure}
    \centering
    \includegraphics[keepaspectratio=true,width=\columnwidth]{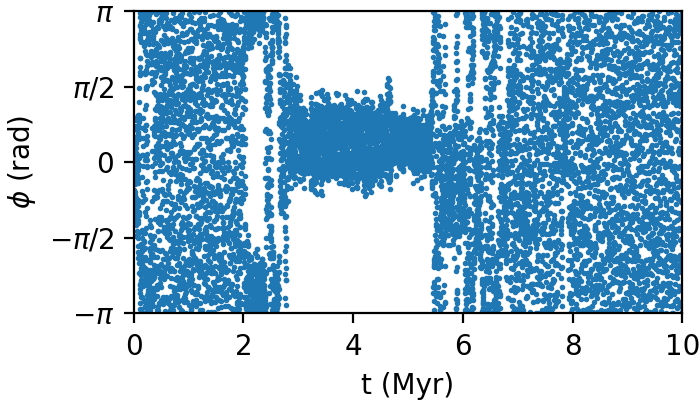}
    \caption{The 4-body resonance angle over time of a perturbed non-resonant \hr system that is near MMR. An MMR chain is formed after approximately 3 Myr and lost after approximately 5.5 Myr as the resonance angle librates during this range of time.}
    \label{fig:oor_near_resonance_hr_4br}
\end{figure}

\subsubsection{\hr}
\label{survival_time_cluster_hr}
\begin{figure*}
    \centering
    \includegraphics[keepaspectratio=true,width=\textwidth]{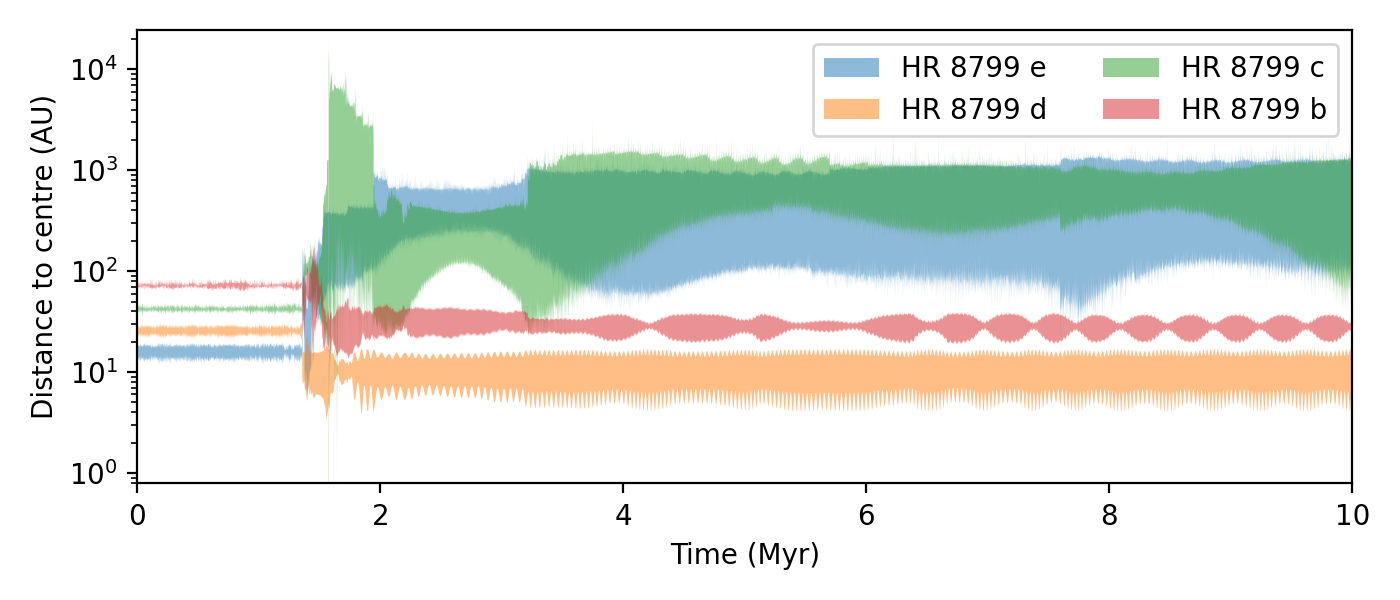}
    \caption{A non-resonant \hr system that dissolved after approximately 0.65 Myr in isolation, but reached the cutoff time of 10 Myr when perturbed by neighbouring cluster stars. The shaded areas show the range between a planet's pericentre and apocentre. After around 1.4 Myr, the system alters its architecture by swapping planets, growing the orbits of \hr e and \hr c, and shrinking the orbit of \hr d. \hr e and \hr c do not interact despite overlapping areas because their orbits are at different orientations.}
    \label{fig:hr_p_sys_24}
\end{figure*}

%82% in MMR survives in cluster vs 100% in MMR in isolation, two out of MMR survive in cluster vs none out of MMR in isolation, one that barely didn't survive in isolation, one that doesn't suddenly get destroyed anymore
The survival rates over time for perturbed \hr systems are shown in figure \ref{fig:survival_rate_hr}. 41 of the 50 resonant systems survive the perturbations from their neighbouring stars, compared to the system surviving in isolation. Curiously, two non-resonant systems now survive until the cutoff time of 10 Myr when perturbed, while none of the non-resonant systems did in isolation. These two systems are the system from section \ref{survival_time_isolation_hr} that almost survived until the cutoff time in isolation, and another system that no longer suddenly dissolves. The first system briefly forms an MMR chain when perturbed by neighbours (see figure \ref{fig:oor_near_resonance_hr_4br}). The second system survives roughly 0.65 Myr in isolation before violently dissolving in the last 20 kyr. In the perturbed simulation, it drastically changes its architecture (see figure \ref{fig:hr_p_sys_24}). Similar to the first system, the second system starts near MMR without an MMR chain, but never forms an MMR chain in either simulation and eventually loses its 2-body MMRs in the first 2 Myr of the perturbed evolution. Figure \ref{fig:survival_rate_hr} shows that not only these two systems survive longer, but that all systems appear to systematically survive longer. The significance of this observation is discussed in appendix \ref{appendix_significance_perturbed_extension} and the required changes to the architecture of the planetary system to explain this observation are further investigated in appendix \ref{appendix_nonresonant_architectural_shifts}.

%11 out of 50 lose MMR, one near end and was omitted, one loses MMR early and restores MMR intermittently and survives until the end and is assumed to survive 10 Myr, nine don't survive, 39 keep MMR and survive and ocassionally increase libration magnitude
The fifth plotted curve of figure \ref{fig:survival_rate_hr} shows the 11 resonant systems that lost resonance, except for one. One of these 11 systems is omitted because it lost resonance less than 100 kyr before the end of the simulation and managed to survive until the cutoff time. Therefore, the exact survival time after losing resonance is unknown, which may incorrectly skew the fifth curve with an underestimated survival time for this system. 9 of the remaining 10 systems that lost resonance do not survive until the end of the simulation. The last system that lost resonance loses resonance after 1.2 Myr and manages to intermittently restore its MMR chain. We cut off the fifth curve at 8.8 Myr as we do not know whether the system survives the last 1.2 Myr. The other 39 resonant systems maintain their MMR chains and occasionally increase their libration magnitudes in their 4-body resonance angles. The resonant systems that lost resonance survive longer from the moment they lose resonance than the perturbed non-resonant systems, as seen by the fifth curve compared to the curve of the perturbed non-resonant systems. This implies that having lost resonance is not as harmful to these systems as being out of resonance due to randomised mean anomalies.

\begin{figure}
    \centering
    \includegraphics[keepaspectratio=true,width=\columnwidth]{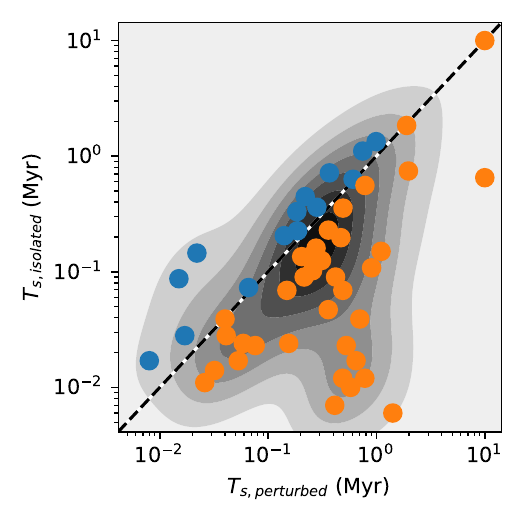}
    \caption{The survival times of the isolated non-resonant systems ($T_{s,isolated}$) versus the survival times of the perturbed non-resonant systems ($T_{s,perturbed}$) for \hr. The systems are biased towards surviving longer when perturbed. The colour of a data point indicates if a system survived longer when isolated (blue, i.e. left of $y=x$) or survived longer when perturbed (orange, i.e. right of $y=x$). The background shows a kernel density estimation.}
    \label{fig:survival_time_nonresonant_comparison_hr}
\end{figure}

%36 out of MMR survive longer in cluster vs 14 that survive shorter
The non-resonant systems survive longer when perturbed by their neighbours; 36 systems survived longer compared to 14 systems that survived shorter (see figure \ref{fig:survival_time_nonresonant_comparison_hr}). This asymmetry between longer and shorter survival times tentatively suggests that being perturbed by neighbours may be beneficial to the survival time of these non-resonant planetary systems.

\subsubsection{\pds}
\begin{figure}
    \centering
    \includegraphics[keepaspectratio=true,width=\columnwidth]{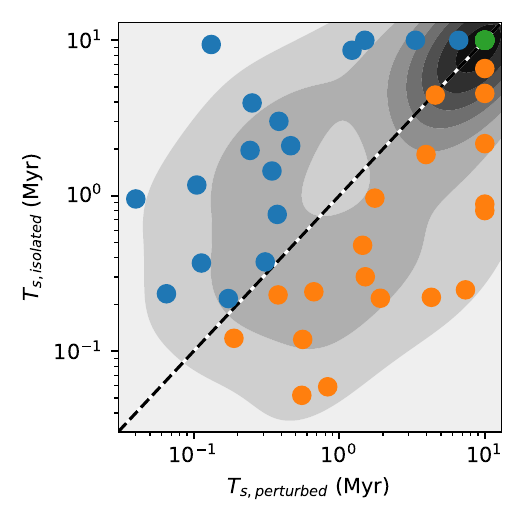}
    \caption{The survival times of the isolated non-resonant systems ($T_{s,isolated}$) versus the survival times of the perturbed non-resonant systems ($T_{s,perturbed}$) for \pds. The systems are not biased towards surviving longer when either isolated or perturbed. The colour of a data point indicates if a system survived longer when isolated (blue, i.e. left of $y=x$), survived longer when perturbed (orange, i.e. right of $y=x$) or survived as long when isolated as when perturbed (green, i.e. on $y=x$). The background shows a kernel density estimation.}
    \label{fig:survival_time_nonresonant_comparison_pds}
\end{figure}

%94% in MMR in cluster vs 100% in MMR in isolation, 19 out of MMR in cluster vs 17 out of MMR in isolation, individuals still strongly affected, 14 survive both in cluster and in isolation, 15 near MMR
The survival rates over time for perturbed \pds systems are shown in figure \ref{fig:survival_rate_pds}. 47 of the 50 resonant systems survive the perturbations from their neighbouring stars, compared to the single system surviving in isolation. The number of perturbed non-resonant systems that survive is nearly identical to the isolated non-resonant systems: 19 perturbed non-resonant systems survive until the cutoff time of 10 Myr versus 17 non-resonant systems doing so in isolation, with an overlap of 14 systems. This implies that the non-resonant \pds systems are not strongly influenced by the perturbations of their neighbours. However, figure \ref{fig:survival_time_nonresonant_comparison_pds} shows that individual survival times can be strongly affected. 15 of the 19 surviving perturbed non-resonant systems start near MMR.

%Out of MMR in cluster: 19 longer vs 17 shorter
19 non-resonant \pds systems survived longer when perturbed versus 17 non-resonant \pds systems that survived longer when isolated, ignoring the 14 non-resonant \pds systems that survived until the cutoff time of 10 Myr both when isolated and when perturbed. This is in contrast to the non-resonant \hr systems, which appear to favour a perturbed environment. Perturbations from neighbouring stars may help stabilise a broken MMR chain and these perturbations may not do so for just a broken MMR.

\section{Discussion}
\label{discussion}

%Intro
The data in figures \ref{fig:survival_rate_hr} and \ref{fig:survival_rate_pds} shows that different observations-conformant configurations of \hr and \pds in different environments have different survival times. Further analysis of this data may impose constraints on the architecture of \hr or \pds. We model the dissolution of the systems as exponential decay using the exponential distribution: $p(t)=\lambda e^{-\lambda t}$ for $t\geq0$. This distribution yields the dissolution rate model $f_d(t)=1-e^{-\lambda t}$, which is equivalent to the survival rate model $f_s(t)=e^{-\lambda t}$. Propagation of errors yields a standard deviation of $\sigma_{f_d}=\sigma_{f_s}=f_s(t)(t\sigma_\lambda+\lambda\sigma_t)$. We omit the isolated resonant systems in this analysis as we only have one data point available for each planetary system. For $n$ dissolution events at times $t_i$, the parameter $\lambda$ is estimated by the inverse of the average of the times $t_i$ \citep{Ross2009}:
\begin{equation}
\label{eq:lambda}
    \lambda\approx\frac{n}{\sum_i t_i},
\end{equation}
with a 1-$\sigma$ confidence interval estimated by the following:
\begin{equation}
\label{eq:lambda_bounds}
    \frac{\chi_{-\sigma,2n}^2}{2\sum_i t_i} < \lambda < \frac{\chi_{+\sigma,2n}^2}{2\sum_i t_i},
\end{equation}
where $\chi_{-\sigma,2n}^2$ and $\chi_{+\sigma,2n}^2$ are the values of a $\chi^2$ distribution with $2n$ degrees of freedom where the distribution's CDF is equal to that of a normal distribution at a value of $-1$ and $+1$, respectively.

However, this estimate of $\lambda$ assumes that all data points are independently drawn from the exponential distribution (i.e. all systems have dissolved), but our data is cut off at a time $T_c$. The set of dissolved systems is effectively drawn from a cut-off exponential distribution $p_c(t)$:
\begin{equation}
    p_c(t)=\begin{cases}
        \frac{\lambda}{1-e^{-\lambda T_c}}e^{-\lambda t} & 0 \leq t \leq T_c \\
        0 & \text{otherwise}
    \end{cases}.
\end{equation}
When applying equation \ref{eq:lambda}, this yields the following observed expected $\lambda_{obs}$ in terms of the true $\lambda$:
\begin{equation}
\label{eq:lambda_corr}
    \frac{1}{\lambda_{obs}}=\overline{t_{obs}}=\int_0^{T_c}p_c(t)tdt=\frac{1}{\lambda}-\frac{T_c}{e^{\lambda T_c}-1},
\end{equation}
which cannot be solved analytically for $\lambda$. Consequently, the observed bounds from equation \ref{eq:lambda_bounds} will also differ from the true bounds (interpreted as a standard deviation $\sigma_{l/u}=|\lambda - \Delta\lambda_{l/u}|$ for a lower or upper bound $\Delta\lambda_{l/u}$):
\begin{equation}
\label{eq:sigma_lambda_corr}
    \sigma=\frac{\left|\frac{\partial}{\partial \lambda_{obs}}\left(\frac{1}{\lambda_{obs}}\right)\right|}{\left|\frac{\partial}{\partial\lambda}\left(\frac{1}{\lambda}-\frac{T_c}{e^{\lambda T_c}-1}\right)\right|}\sigma_{obs}=\frac{\sigma_{obs}}{\lambda_{obs}^2\left|\frac{T_c^2e^{\lambda T_c}}{\left(e^{\lambda T_c}-1\right)^2}-\frac{1}{\lambda^2}\right|}.
\end{equation}
We obtain corrected values for $\lambda$ by solving equation \ref{eq:lambda_corr} numerically and obtain corrected uncertainties for $\lambda$ by using equation \ref{eq:sigma_lambda_corr}.

Finally, the expected survival fraction at the age of a system $T_0$ is $f_s(T_0)=e^{-\lambda T_0}$, which is better suited in terms of the natural logarithm when analysing its error: $\ln{f_s(T_0)}=-\lambda T_0$. The error on the natural logarithm of this fraction by propagation of errors is $\sigma_{\ln{f_s}}(T_0)=\sqrt{T_0^2\sigma_\lambda^2+\lambda^2\sigma_{T_0}^2}$ for a normally distributed age and $\sigma_{\ln{f_s}}(T_0)=T_0\sigma_\lambda$ for a boundary age.

\begin{table*}
    \centering
    \begin{tabular}{| l | l | c | c | c |}
        \hline
        Planetary system (age) & Configuration & $\lambda$ (Myr\textsuperscript{-1}) & $\ln{f_s}$ & $f_s$ (5-$\sigma$ CI) \\
        \hline
        & Resonant, perturbed & $0.17\pm0.16$ & $-3.4\pm3.3$ & $(2.4\cdot10^{-9}, 1]$ \\
        \hr ($20$ Myr) & Non-resonant, isolated & $2.29\pm0.32$ & $-46\pm6.5$ & $(1.2\cdot10^{-34}, 1.3\cdot10^{-6})$ \\
        & Non-resonant, perturbed & $2.31\pm0.33$ & $-46\pm6.6$ & $(3.7\cdot10^{-35}, 2.4\cdot10^{-6})$ \\
        \hline
        & Resonant, perturbed & $0.17\pm0.16$ & $-8.6\pm8.2$ & $(2.7\cdot10^{-22}, 1]$ \\
        \hr ($50$ Myr) & Non-resonant, isolated & $2.29\pm0.32$ & $-114\pm16$ & $(1.7\cdot10^{-85}, 2.1\cdot10^{-15})$ \\
        & Non-resonant, perturbed & $2.31\pm0.33$ & $-115\pm17$ & $(8.1\cdot10^{-87}, 8.7\cdot10^{-15})$ \\
        \hline
        & Resonant, perturbed & $0.52\pm0.32$ & $-2.8\pm1.8$ & $(7.1\cdot10^{-6}, 1]$ \\
        \pds ($5.4\pm1.0$ Myr) & Non-resonant, isolated & $0.55\pm0.11$ & $-2.9\pm0.8$ & $(1.0\cdot10^{-3}, 1]$ \\
        & Non-resonant, perturbed & $0.67\pm0.13$ & $-3.6\pm1.0$ & $(2.2\cdot10^{-4}, 1]$ \\
        \hline
    \end{tabular}
    \caption{The exponential distribution parameter $\lambda$, the natural logarithm of the expected fraction of surviving planetary systems at the current age $\ln(f_s)$, and the 5-$\sigma$ confidence interval of the expected fraction of surviving planetary systems at the current age $f_s$ (bounded by $[0,1]$) for a $20$ Myr old \hr, a $50$ Myr old \hr, and a $5.4\pm1.0$ Myr old \pds in the configurations in resonance and perturbed, out of resonance and isolated, and out of resonance and perturbed.}
    \label{table:survival_stats}
\end{table*}

$\lambda$, $\ln{f_s}$ and a 5-$\sigma$ confidence interval (CI) of $f_s$ of \hr (for a boundary age of both $20$ Myr and $50$ Myr) and \pds can be found in table \ref{table:survival_stats}. The CIs for \hr show that \hr's planetary system is only expected to survive to its current age if and only if it is in an MMR chain as the upper bound of the expected survival fraction is negligible for an \hr planetary system that is out of resonance. Perturbed resonant \hr systems can dissolve due to interactions with neighbouring stars, but we cannot rule out with complete confidence that \hr would not survive such an environment as the upper bound of the expected survival fraction is not negligible for a perturbed \hr planetary system that is in resonance. There exists a little overlap between the resonant and non-resonant CIs. The value for $\ln{f_s}$ for the perturbed resonant systems encounters the non-resonant CIs after $3.1 \sigma$ and $2.9 \sigma$ for the isolated and perturbed CIs, respectively, and the non-resonant CIs encounter the perturbed resonant CIs after $4.0\sigma$ in both cases. From these $\sigma$ values, the probability that both true values are in the other's CI is $5.4\sigma$ between the perturbed resonant CI and the isolated non-resonant CI, and $5.3\sigma$ between the perturbed resonant CI and the perturbed non-resonant CI. Therefore, we believe the overlap between the CIs should not be an issue.

While we expect the resonant \pds systems to dissolve less than the non-resonant \pds systems as the means of $\ln{f_s}$ prefer the perturbed resonant systems, we cannot with confidence claim anything about the presence of MMR in \pds or the environment of \pds as the upper bounds of all three CIs of the expected survival fraction are not negligible. It is also clear that there is much overlap between the resonant and non-resonant CIs as the values for $\ln{f_s}$, being within $1 \sigma$ of each other, are statistically indistinguishable.

\section{Conclusion}
\label{conclusion}
%Conclusions: HR has MMR chain, PDS potentially in MMR, unused: MMR chain trades increased susceptibility to external perturbations for increased internal stability, planetary systems without MMR can survive longer in a cluster than in isolation
We have performed numerical simulations of the planetary systems \hr and \pds to investigate the existence of a mean-motion resonance in \hr and the effects of perturbations from neighbouring cluster stars on planetary systems in mean-motion resonance, with a mean-motion resonance chain, with a broken mean-motion resonance, and with a broken mean-motion resonance chain. We summarise our findings as follows:
\begin{itemize}
    \item The planetary system \hr must have a mean-motion resonance chain for the system to have survived in its current state to its present age of 20 Myr to 50 Myr. We expect an \hr system without a mean-motion resonance chain to have dissolved before its present age is reached. We argue that it is less probable that \hr still resides in a cluster environment, but we cannot rule this out with absolute certainty.
    \item We argue that it is more probable that the planetary system \pds is in mean-motion resonance for the system to have survived to its present age of $5.4\pm1.0$ Myr, though we cannot rule out the absence of mean-motion resonance. Similarly, we argue that it is less probable that \pds still resides in a cluster environment, but we cannot rule this out with absolute certainty.
    \item It is almost certainly possible for systems with a near-resonant or broken mean-motion resonance chain to survive longer when perturbed by stellar encounters compared to isolation. Specifically, we observe an average increase in survival time of approximately 173\% when perturbed by stellar encounters for \hr systems with a broken mean-motion resonance chain.
\end{itemize}

%Perturbed and isolated compared to stability literature
The results of the isolated non-resonant systems agree with previous works on planetary system stability: both the \hr systems and the \pds systems violate their respective stability criteria from \citet{Gladman1993}, \citet{Chambers1996} and \citet{Smith2009}, and both these systems steadily dissolve when in isolation and out of mean-motion resonance. When compared to the isolated non-resonant results, the perturbed non-resonant \hr systems are almost certainly more stable than their isolated versions, which would result in a slightly different limit to their stability criterion. However, the perturbed non-resonant \hr systems still dissolve at a comparable rate as the isolated non-resonant \hr systems and this difference is not observed in the non-resonant \pds systems. This difference is far clearer when comparing non-resonant and resonant systems: despite the stability criteria treating the resonant and non-resonant systems equally, the resonant systems are far more stable than the non-resonant systems. This indicates that the stability criteria may not be accurate using masses, orbital elements and number of planets alone, and that an additional term for mean-motion resonances should be included, or that a separate class of stability criteria is needed for planetary systems in mean-motion resonance.

\begin{acknowledgements}
%TODO: Mail ALICE after publication: https://pubappslu.atlassian.net/wiki/spaces/HPCWIKI/pages/37519428/Acknowledging+ALICE
This work was performed using the compute resources from the Academic Leiden Interdisciplinary Cluster Environment (ALICE) provided by Leiden University.

The testing and simulation jobs for this work used computing resources for a total of 125 days, 11 hours, 29 minutes and 37 seconds. On average and rounded up, these jobs used 14 cores on a Xeon Gold 6126 and 23 GB of memory. These jobs spent a total of 705 kWh of energy, which is equivalent to 264 kg of CO\textsubscript{2} emission in the Netherlands, according to \url{https://calculator.green-algorithms.org} (approximating using a Xeon Gold 6142).

This research has made use of the VizieR catalogue access tool, CDS, Strasbourg, France (DOI : 10.26093/cds/vizier). The original description of the VizieR service was published in 2000, A\&AS 143, 23.

This work has made use of data from the European Space Agency (ESA) mission {\it Gaia} (\url{https://www.cosmos.esa.int/gaia}), processed by the {\it Gaia} Data Processing and Analysis Consortium (DPAC, \url{https://www.cosmos.esa.int/web/gaia/dpac/consortium}). Funding for the DPAC has been provided by national institutions, in particular the institutions participating in the {\it Gaia} Multilateral Agreement.

This work used the Python programming language \citep{python} and the Python packages NumPy \citep{harris2020array}, SciPy \citep{2020SciPy-NMeth}, Matplotlib \citep{Hunter2007}, AMUSE \citep{PortegiesZwart2008,PortegiesZwart2013,Pelupessy2013,portegieszwart2018,amuselatest} and h5py \citep{collette_python_hdf5_2014}.
\end{acknowledgements}

\bibliographystyle{aa}
\bibliography{paper1.bib}

\begin{appendix}

\section{Planetary system initial conditions}
\label{appendix_initial_conditions}

\begin{table}[h]
    \centering
    \begin{tabular}{| l | c | c | c | c | c |}
        \hline
        Planet & $m$ ($M_\text{J}$) & $a$ (AU) & $e$ & $\nu$ & $\omega$ \\
        \hline
        \hr e & 7.4 & 15.8 & 0.14 & 1.88\textsuperscript{\textdagger} & 4.67 \\
        \hr d & 9.1 & 26.2 & 0.12 & 5.87\textsuperscript{\textdagger} & 6.23 \\
        \hr c & 7.8 & 40.7 & 0.04 & 3.85\textsuperscript{\textdagger} & 5.26 \\
        \hr b & 5.7 & 73.3 & 0.04 & 0.14\textsuperscript{\textdagger} & 0.50 \\
        \hline
        \pds b & 3.2 & 20.8 & 0.19 & 4.16\textsuperscript{\textdagger} & 3.60 \\
        \pds c & 7.5 & 34.3 & 0.03 & 5.52\textsuperscript{\textdagger} & 3.33 \\
        \hline
    \end{tabular}
    \caption{The mass $M$ in Jupiter masses and the orbital parameters $a$ (semimajor axis), $e$ (eccentricity), $\nu$ (true anomaly) and $\omega$ (argument of pericentre) of the resonant initial conditions of \hr and \pds. \diffadd{The non-resonant initial conditions are equal to the resonant initial conditions, except for the values marked with \textdagger, which are drawn from a uniform random distribution between $0$ and $2\pi$ for each unique copy of initial conditions.} All orbits are planar, so all inclinations $i = 0$ and all longitudes of ascending node $\Omega = 0$.}
    \label{table:initial_conditions}
\end{table}

\begin{diffenable}
\begin{table}[h]
    \centering
    \begin{tabular}{| l | c | c | c | c | c |}
        \hline
        Planet & $m$ ($M_\text{J}$) & $a$ (AU) & $e$ & $\nu$ & $\omega$ \\
        \hline
        \hr e & \textdagger & \textdagger & \textdagger & $U(0,2\pi)$ & \textdagger \\
        \hr d & \textdagger & \textdagger & \textdagger & $U(0,2\pi)$ & \textdagger \\
        \hr c & \textdagger & \textdagger & \textdagger & $U(0,2\pi)$ & \textdagger \\
        \hr b & \textdagger & \textdagger & \textdagger & $U(0,2\pi)$ & \textdagger \\
        \hline
        \pds b & \textdagger & \textdagger & \textdagger & $U(0,2\pi)$ & \textdagger \\
        \pds c & \textdagger & \textdagger & \textdagger & $U(0,2\pi)$ & \textdagger \\
        \hline
    \end{tabular}
    \caption{\diffrem{The mass $M$ in Jupiter masses and the orbital parameters $a$ (semimajor axis), $e$ (eccentricity), $\nu$ (true anomaly) and $\omega$ (argument of pericentre) of the non-resonant initial conditions of \hr and \pds. All orbits are planar, so all inclinations $i = 0$ and all longitudes of ascending node $\Omega = 0$. Values marked with {\textdagger} are equal to the value listed in table \ref{table:initial_conditions}. Values marked with $U(0,2\pi)$ are drawn from a uniform random distribution between $0$ and $2\pi$ for each unique copy of initial conditions.}}
    \label{table:initial_conditions_nonresonant}
\end{table}
\end{diffenable}

\section{Resonance angles}
\label{resonance_angles}
%Resonance angles
For a given $j-o_i:j$ MMR of order $o_i$ between two bodies, we define a 2-body resonance angle using their mean anomalies $\lambda_i$ and an argument of pericentre $\overline{\omega}_{i+X}$, where $X\in\{0,1\}$ indicates whether we use the inner ($X=0$) or outer ($X=1$) argument of pericentre \citep{Huang2021}:
\begin{equation}
    \phi_{2BR,i,X}=(j-o_i)\lambda_i-j\lambda_{i+1}+o_i\overline{\omega}_{i+X}.
\end{equation}
For \pds, we can simplify this equation because we know its only MMR is 2:1 ($j=2$, $o_i=1$). We can also translate the indices $0$ and $1$ to the planet labels $b$ and $c$:
\begin{equation}
    \phi_{2BR,PDS 70,X}=\lambda_b-2\lambda_c+\overline{\omega}_{X}.
\end{equation}
We can also simplify the equation for \hr since its resonances are all 2:1:
\begin{equation}
    \phi_{2BR,HR 8799,i,X}=\lambda_i-2\lambda_{i+1}+\overline{\omega}_{i+X}.
\end{equation}
We can obtain a 3-body resonance angle for \hr by solving for $\overline{\omega}_{i+1}$ in the cases of $i=i;X=1$ and $i=i+1;X=0$ \citep{Huang2021}:
We can combine the 2-body resonance angles of $i=i;X=1$ and of $i=i+1;X=0$ and solve for $\overline{\omega}_{i+1}$ to obtain a 3-body resonance angle for \hr \citep{Huang2021}:
\begin{equation}
    \phi_{3BR,HR 8799,i}=\lambda_i-3\lambda_{i+1}+2\lambda_{i+2}.
\end{equation}
Finally, we can sum the 3-body resonance angles of $i=0$ and of $i=1$ to get a 4-body resonance angle that describes the entire planetary system of \hr \citep{GozdziewskiMigaszewski2020}, and translate the indices $0$, $1$, $2$ and $3$ to the planet labels $e$, $d$, $c$ and $b$:
\begin{equation}
    \phi_{4BR,HR 8799}=\lambda_e-2\lambda_d-\lambda_c+2\lambda_b.
\end{equation}

\section{Extended survival of perturbed broken MMR chains}
\label{appendix_significance_perturbed_extension}
%Surviving longer in cluster?
From our simulations, it appears that planetary systems with a broken MMR chain (the non-resonant \hr systems) survive longer when perturbed by neighbouring cluster stars, whereas planetary systems with a broken MMR (the non-resonant \pds systems) are negligibly affected by these perturbations. This suggests that perturbing a broken MMR chain somehow allows a planetary system to survive longer.

%One-sided binomial test
The data of \hr (see figure \ref{fig:survival_time_nonresonant_comparison_hr}) does not contain definitive survival times for each non-resonant planetary system as 2 of the 50 perturbed survived until the cutoff time of 10 Myr. However, we can tell whether a system survived longer or not since none of the isolated non-resonant systems survived until the cutoff time. This means we have a complete dataset to perform a one-sided binomial test. We have 36 out of 50 systems surviving longer when perturbed than when isolated. For the null hypothesis, we assume that the survival time is unaffected by the presence of perturbing neighbours, and therefore the probabilities of surviving longer and shorter are equal; $H_0: P=0.5$. This one-sided binomial test yields a $p$-value of $0.0013$ for \hr, which is equivalent to a $3.22$-$\sigma$ result for a standard two-sided normal distribution. The data of \pds (see figure \ref{fig:survival_time_nonresonant_comparison_pds}) is not even complete for this one-sided binomial test as 14 non-resonant systems survived until the cutoff time both when isolated and when perturbed. These 14 systems do not provide any information about whether perturbing neighbours affect the survival time of a system, therefore we omit them so we can perform a one-sided binomial test. We are left with 36 non-resonant systems, of which 19 survived longer when being perturbed by neighbouring stars. This results in a $p$-value of $0.43$, which is equivalent to $0.78$-$\sigma$.

%Alternative statistical test in log space
Alternatively, we can do a more quantitative analysis anyway. We do this by asserting that the survival time of a non-resonant system that has reached the cutoff time is best estimated by the cutoff time. If the survival time is unaffected by perturbing neighbours, then the data in figures \ref{fig:survival_time_nonresonant_comparison_hr} and \ref{fig:survival_time_nonresonant_comparison_pds} should be evenly spread around $y=x$. On the other hand, if these perturbations do affect the survival time, the data should be evenly spread around $y=\alpha x$, where $\alpha$ is greater or smaller than 1, depending on whether isolated systems or perturbed systems survive longer. Each data point in the figure provides a $\alpha_i$ which is equal to $T_{s,isolated,i}/T_{s,perturbed,i}$. However, these $\alpha$'s do not provide a symmetric space when mirroring over $y=x$ (e.g. $y=3x$ and $y=x/3$ yield $\alpha=3$ and $\alpha=1/3$). Instead, we use $\beta=\ln{\alpha}$ and $\beta_i=\ln{\alpha}$ to remedy this issue. We can estimate $\beta$ by taking the mean and standard error of the individual $\beta_i$'s from our data. This yields $\beta=-1.01\pm0.23$ for \hr, equivalent to a $4.45$-$\sigma$ deviation from $y=x$ in favour of perturbed systems surviving longer, and $\beta=-0.0045\pm0.23$ for \pds, equivalent to a $0.020$-$\sigma$ deviation from $y=x$ in favour of perturbed systems surviving longer. We know that the result for \hr is slightly underestimated as only two perturbed non-resonant systems survived until the cutoff time and no isolated non-resonant systems did. It is impossible to know if \pds's result is underestimated or overestimated since a notable fraction of both isolated and perturbed non-resonant systems survived until the cutoff time. If we translate the resulting $\beta$ values back to $\alpha$'s, we get that we expect non-resonant \hr systems to survive approximately $173\%$ longer on average and non-resonant \pds systems to survive approximately $0.45\%$ longer on average when either is perturbed by neighbouring stars.

%Test comparison + conclusion
For \hr, the one-sided binomial test yielded the weaker of the two results. This is not surprising since it treats each data point as equal, while figure \ref{fig:survival_time_nonresonant_comparison_hr} shows that the systems that survived longer when perturbed produce more extreme deviations from $y=x$ than the systems that survived longer in isolation. This means the one-sided binomial test underestimates the effect the perturbations have on the survival times of the systems. This issue is addressed by the more quantitative estimation of $\beta$, though at a minor cost to data completeness. The data completeness cost for \hr is small as only 2 out of 50 systems do not provide complete data and we know that the result is slightly underestimated because of it. Both tests for \hr agree, so we conclude that wide planetary systems like \hr with a broken MMR chain survive longer when perturbed by neighbouring cluster stars than when isolated. The $4.45$-$\sigma$ result from the estimation of $\beta$ provides fairly accurate, though conservative statistical evidence for this conclusion. For \pds, neither result is statistically significant, so we can only conclude that there is no indication that the survival times of wide planetary systems like \pds with a broken MMR are affected by perturbations of neighbouring cluster stars.

\section{Non-resonant architectural shifts}
\label{appendix_nonresonant_architectural_shifts}
%What kind of architecture shifts causes planetary systems to survive longer?
As seen in section \ref{survival_time_cluster_hr}, perturbations by neighbouring cluster stars may cause a non-resonant \hr system to alter its architecture. This somehow allows the planetary system to survive longer than in isolation. We want to find out what kind of shifts in architecture are caused by the perturbations and how these changes affect the survival time of a planetary system. We analyse the orbital parameters of both the isolated and the perturbed non-resonant planetary systems of both \hr and \pds. Any asymmetry between isolated and perturbed systems or between \hr and \pds might be able to explain the qualitative differences between figure \ref{fig:survival_time_nonresonant_comparison_hr} and figure \ref{fig:survival_time_nonresonant_comparison_pds}.

\subsection{\hr}
\begin{figure*}
    \centering
    \begin{subfigure}[t]{\textwidth}
        \centering
        \includegraphics[keepaspectratio,width=\linewidth]{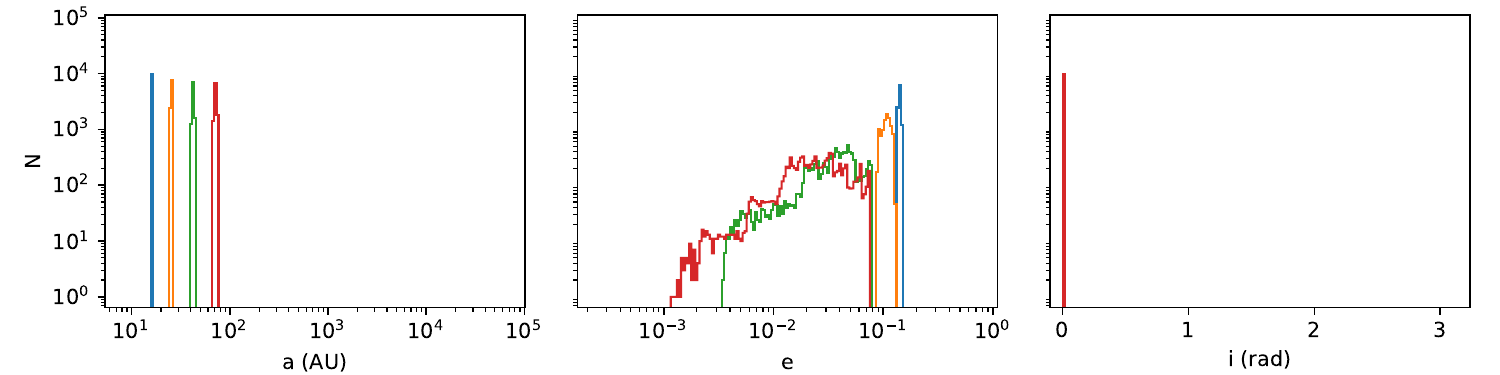}
        \caption{The isolated resonant systems. Orbital evolution oscillates around the initial conditions and the MMR chain is maintained.}
        \label{fig:hr_arch_p_sys_isolated}
    \end{subfigure}
    \newline
    \begin{subfigure}[t]{\textwidth}
        \centering
        \includegraphics[keepaspectratio,width=\linewidth]{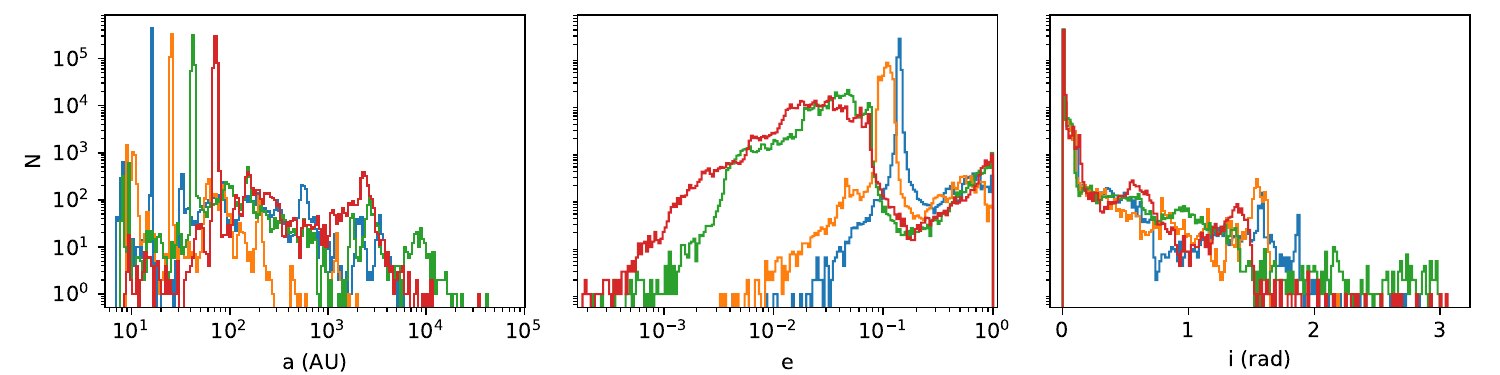}
        \caption{The perturbed resonant systems. The MMR chain is broken in some simulations. However, these broken MMR chains are often short-lived, so the initial MMR chain remains the only stable mode.}
        \label{fig:hr_arch_p_sys}
    \end{subfigure}
    \newline
    \begin{subfigure}[t]{\textwidth}
        \centering
        \includegraphics[keepaspectratio,width=\linewidth]{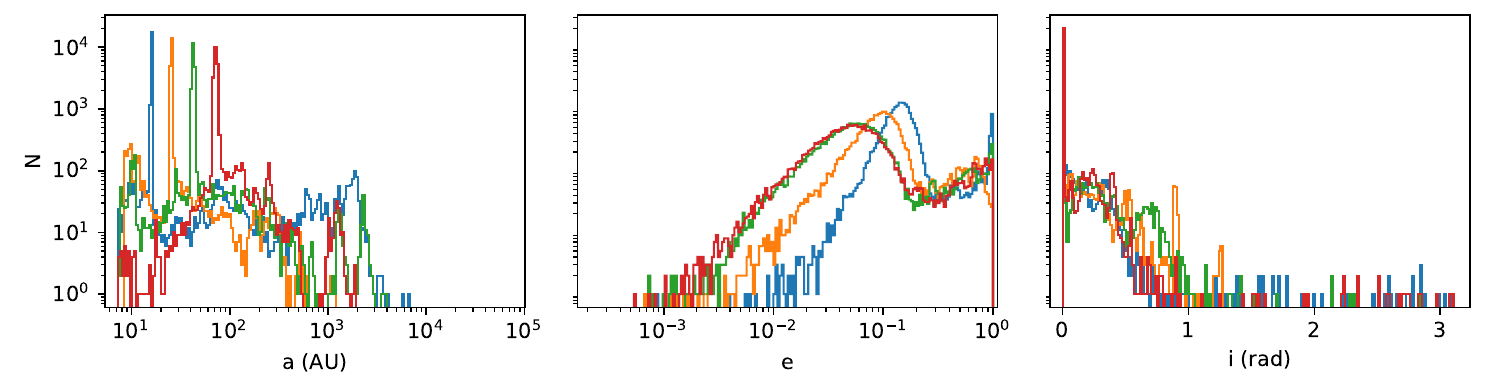}
        \caption{The isolated non-resonant systems. Some of the systems diverge from the broken MMR chain and end up destroyed. The eccentricities are not as sharply bound as the resonant systems. Shifts in inclination occur despite the systems being planar and isolated. This should not happen as there is no possible force to drive the planets off the plane. We believe this is the result of chaotic growth of numerical errors as these systems do not just live in two-dimensional space, but are randomly orientated in three-dimensional space.}
        \label{fig:hr_arch_p_sys_oor_isolated}
    \end{subfigure}
    \newline
    \begin{subfigure}[t]{\textwidth}
        \centering
        \includegraphics[keepaspectratio,width=\linewidth]{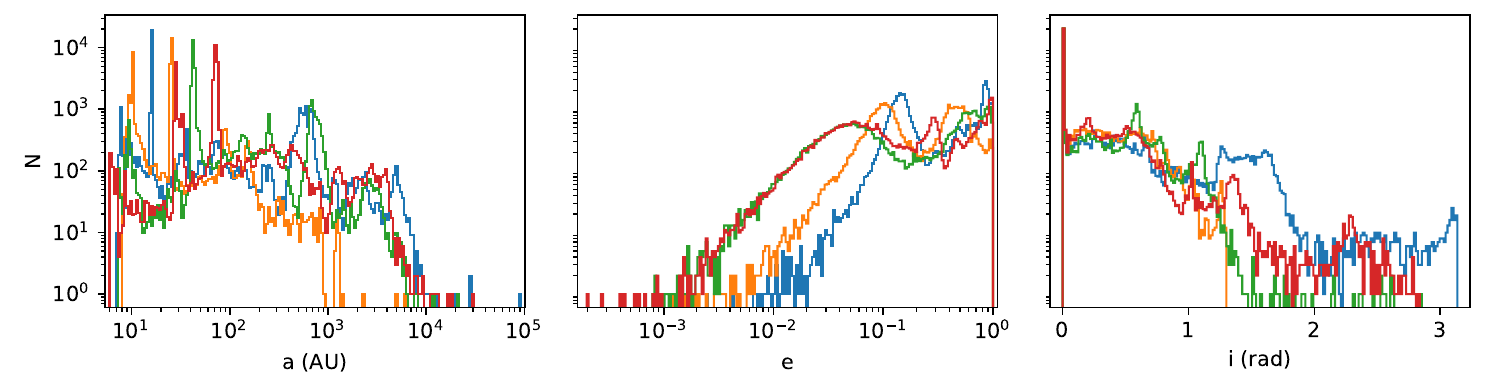}
        \caption{The perturbed non-resonant systems. States that diverged from the broken MMR chain are more prominent than in figures \ref{fig:hr_arch_p_sys} and \ref{fig:hr_arch_p_sys_oor_isolated}. \hr d and \hr b show an apparent second stable mode, but this only occurs in a single simulation. The broken MMR chain remains the most prominent feature.}
        \label{fig:hr_arch_p_sys_oor}
    \end{subfigure}
    \caption{The histograms of the semi-major axes, eccentricities and inclinations of the \hr systems, separated by planet. The dataset of each of the histograms contains all states of all its relevant simulations. These are 10000 states per simulation, with 1 simulation for \ref{fig:hr_arch_p_sys_isolated}, and 50 simulations for \ref{fig:hr_arch_p_sys}, \ref{fig:hr_arch_p_sys_oor_isolated} and \ref{fig:hr_arch_p_sys_oor}. By including all states instead of just the final state, longer-lived states become more prominent even if these states are not the final state.}
    \label{fig:hr_arch}
\end{figure*}

\begin{figure*}
    \centering
    \begin{subfigure}[t]{\textwidth}
        \centering
        \includegraphics[keepaspectratio,width=\linewidth]{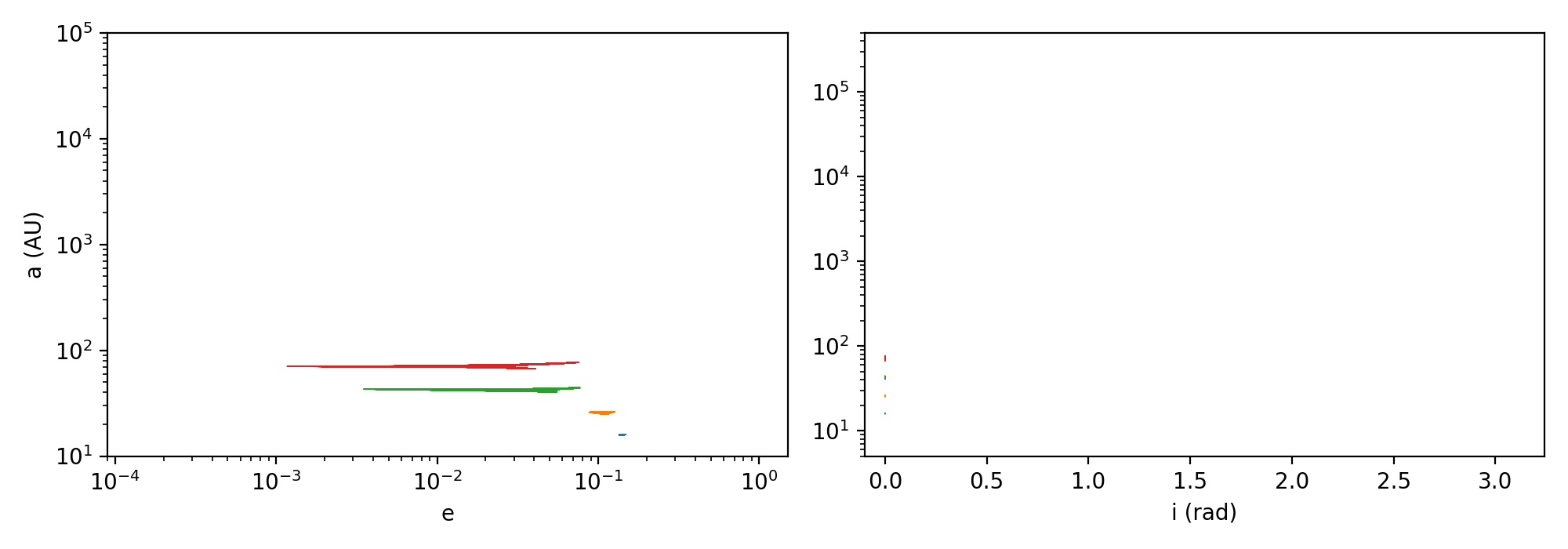}
        \caption{An isolated resonant \hr system. All four planets remain planar and oscillate around their initial semi-major axis and eccentricity.}
        \label{fig:parameters_hr_p_sys_isolated_0}
    \end{subfigure}
    \newline
    \begin{subfigure}[t]{\textwidth}
        \centering
        \includegraphics[keepaspectratio,width=\linewidth]{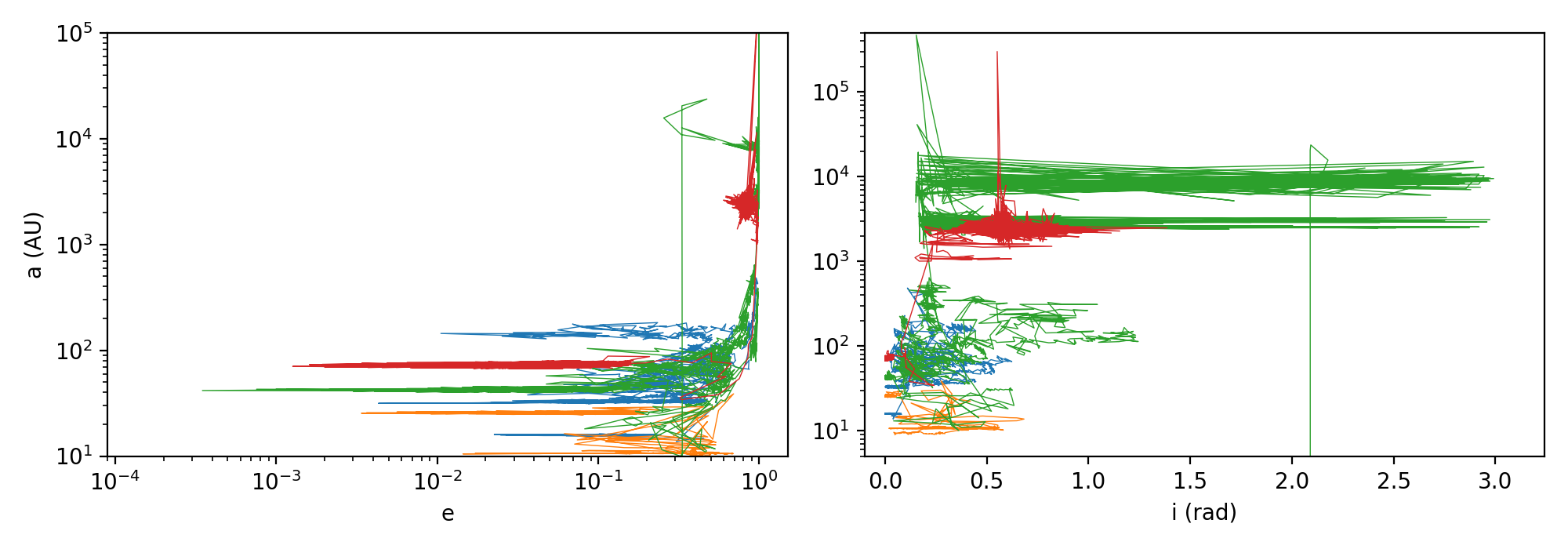}
        \caption{A perturbed resonant \hr system that did not manage to survive until the cutoff time. Only \hr b's semi-major axis finds some stability after a strong perturbation before a planet is lost.}
        \label{fig:parameters_hr_p_sys_33}
    \end{subfigure}
    \newline
    \begin{subfigure}[t]{\textwidth}
        \centering
        \includegraphics[keepaspectratio,width=\linewidth]{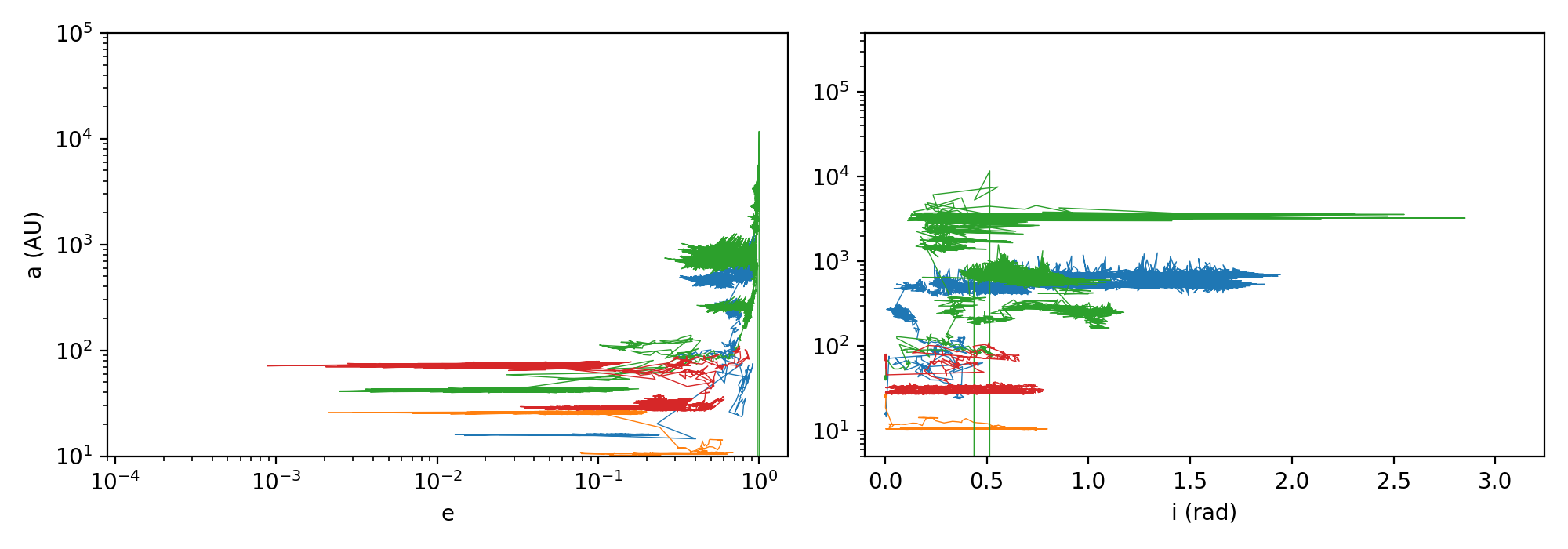}
        \caption{The perturbed non-resonant \hr system that managed to survive until the cutoff time despite dissolving in isolation (see section \ref{survival_time_cluster_hr}). Two planets move inwards and two planets move outwards. All planets, except for \hr c (green), remain at their new semi-major axes, but the orientations of the orbits are not fixed.}
        \label{fig:parameters_hr_p_sys_oor_24}
    \end{subfigure}
    \caption{The paths of semi-major axis versus eccentricity and semi-major axis versus inclination for a representative set of three different \hr systems.}
    \label{fig:parameters_hr}
\end{figure*}

%No evolution in MMR in isolation, negligible evolution in MMR in cluster
As shown in figure \ref{fig:hr_arch_p_sys_isolated} and \ref{fig:parameters_hr_p_sys_isolated_0}, the isolated resonant \hr system does not evolve apart from some oscillation around its initial conditions. When the resonant \hr system gets perturbed by its neighbours (figure \ref{fig:hr_arch_p_sys}), the effects of the perturbations become visible as noise-like deviations from the initial conditions. The semi-major axes are scattered up to approximately 10000 AU, the eccentricities go up to 1, and significant deviations can be found in the inclinations. An example is shown in figure \ref{fig:parameters_hr_p_sys_33}. However, all these deviations are short-lived, and the single stable mode corresponds to the initial conditions.

%Negligible evolution out of MMR in isolation, less negligible and wider evolution and second stable mode from single system out of MMR in cluster
The isolated non-resonant systems in figure \ref{fig:hr_arch_p_sys_oor_isolated} also have the initial conditions as the dominant mode. Like the perturbed resonant systems, the orbital parameters are no longer bound to just the initial conditions as unstable systems get perturbed by self-interaction. However, the semi-major axes only got up to roughly 3000 AU and the inclinations rarely go beyond 1 rad. The perturbed non-resonant systems (see figure \ref{fig:hr_arch_p_sys_oor}) are similar to the perturbed resonant systems, though the noise-like continuum is slightly elevated, which indicates that slightly more systems get perturbed to fill these parts of parameter space. The perturbed non-resonant systems reach the same semi-major axis bound as the perturbed resonant systems of around 10000 AU. Figure \ref{fig:hr_arch_p_sys_oor} also shows a second apparent stable mode for \hr d and \hr b. However, this mode is explained by the single system from section \ref{survival_time_cluster_hr} that adapted into a seemingly stable configuration early in its lifetime (see figure \ref{fig:hr_p_sys_24}), also shown in figure \ref{fig:parameters_hr_p_sys_oor_24}.

%Perturbations produce wider deviations than being out of MMR
Figure \ref{fig:hr_arch} shows that being perturbed by a neighbouring cluster star and being out of resonance can produce a variety of changes in the architecture of the \hr systems and that these changes are rarely stable. The perturbations from neighbouring cluster stars produce wider deviations in semi-major axis than being out of resonance, which can explain the difference between isolated and perturbed non-resonant systems seen in figure \ref{fig:survival_rate_hr} and discussed in appendix \ref{appendix_significance_perturbed_extension}. The wider unstable configurations take longer to complete the necessary orbits to destroy themselves through internal interactions than the less wide unstable configurations. Therefore, the perturbed non-resonant systems that are perturbed up to 10000 AU will manage to survive longer than the isolated non-resonant systems that are perturbed up to 3000 AU.

\subsection{\pds}
\begin{figure*}
    \centering
    \begin{subfigure}[t]{\textwidth}
        \centering
        \includegraphics[keepaspectratio,width=\linewidth]{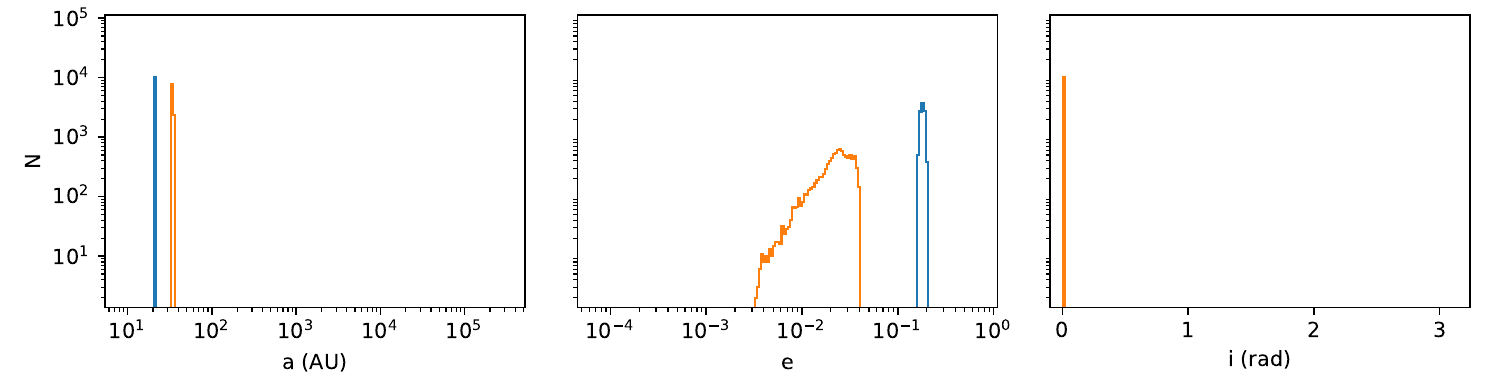}
        \caption{The isolated resonant systems. Orbital evolution oscillates around the initial conditions and the MMR chain is maintained.}
        \label{fig:pds_arch_p_sys_isolated}
    \end{subfigure}
    \newline
    \begin{subfigure}[t]{\textwidth}
        \centering
        \includegraphics[keepaspectratio,width=\linewidth]{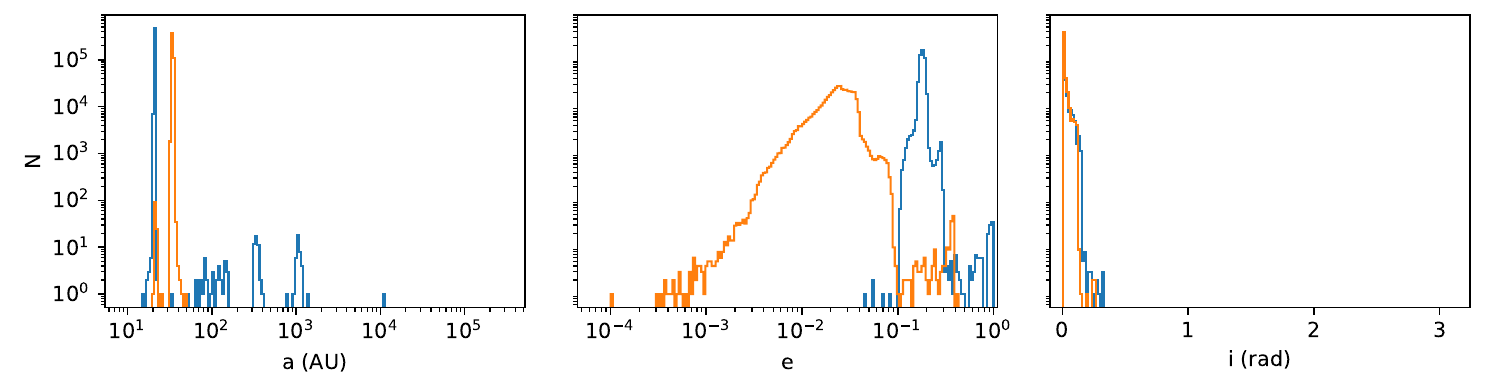}
        \caption{The perturbed resonant systems. The MMR is rarely broken and any systems with a broken MMR are rapidly destroyed. The inclinations do not diverge far from planar.}
        \label{fig:pds_arch_p_sys}
    \end{subfigure}
    \newline
    \begin{subfigure}[t]{\textwidth}
        \centering
        \includegraphics[keepaspectratio,width=\linewidth]{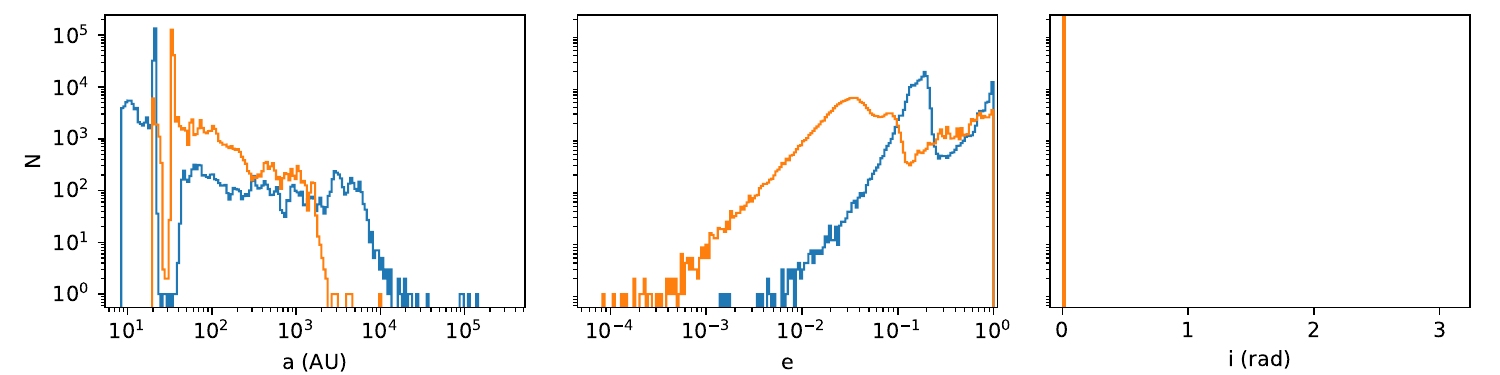}
        \caption{The isolated non-resonant systems. Some of the systems diverge from the broken MMR. A semi-major axis valley is present for \pds b around the initial semi-major axis of \pds c. The eccentricities are not as sharply bound as the resonant systems. Some of the systems get perturbed violently enough that the direction of \pds b reverses. Despite the issues with random orientations in figure \ref{fig:hr_arch_p_sys_oor_isolated}, all systems remain planar here. We believe these systems can manage to stay planar because they contain fewer bodies to encourage the chaotic growth of numerical errors.}
        \label{fig:pds_arch_p_sys_oor_isolated}
    \end{subfigure}
    \newline
    \begin{subfigure}[t]{\textwidth}
        \centering
        \includegraphics[keepaspectratio,width=\linewidth]{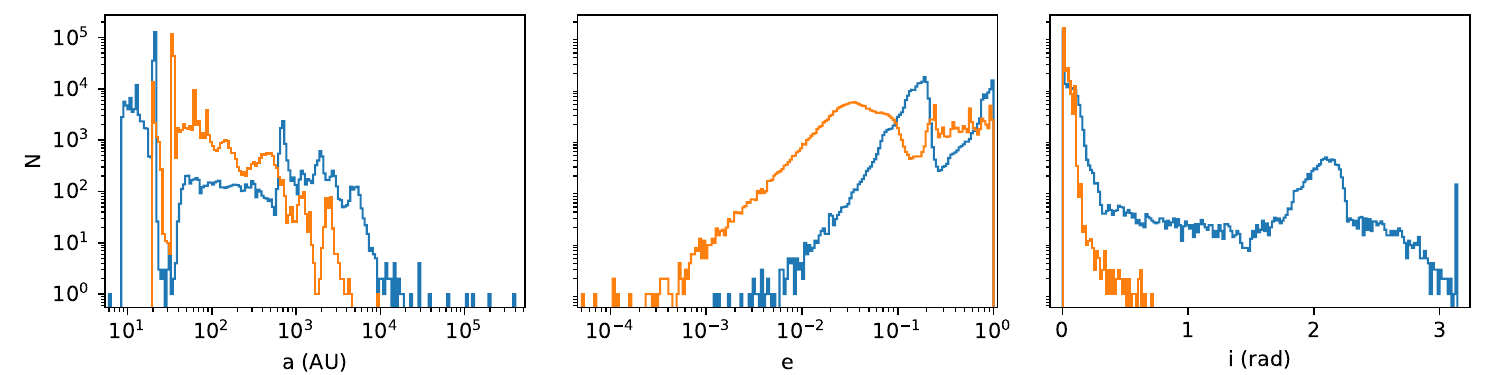}
        \caption{The perturbed non-resonant systems. The semi-major axes and the eccentricities are nearly identical to the semi-major axes and the eccentricities in figure \ref{fig:pds_arch_p_sys_oor_isolated}. In contrast to figures \ref{fig:pds_arch_p_sys} and \ref{fig:pds_arch_p_sys_oor_isolated}, the inclinations are more strongly perturbed and now cover all of the possible parameter space for \pds b.}
        \label{fig:pds_arch_p_sys_oor}
    \end{subfigure}
    \caption{The histograms of the semi-major axes, eccentricities and inclinations of the \pds systems, separated by planet. The dataset of each of the histograms contains all states of all its relevant simulations. These are 10000 states per simulation, with 1 simulation for \ref{fig:pds_arch_p_sys_isolated}, and 50 simulations for \ref{fig:pds_arch_p_sys}, \ref{fig:pds_arch_p_sys_oor_isolated} and \ref{fig:pds_arch_p_sys_oor}. By including all states instead of just the final state, longer-lived states become more prominent even if these states are not the final state.}
    \label{fig:pds_arch}
\end{figure*}

\begin{figure*}
    \centering
    \begin{subfigure}[t]{\textwidth}
        \centering
        \includegraphics[keepaspectratio,width=\linewidth]{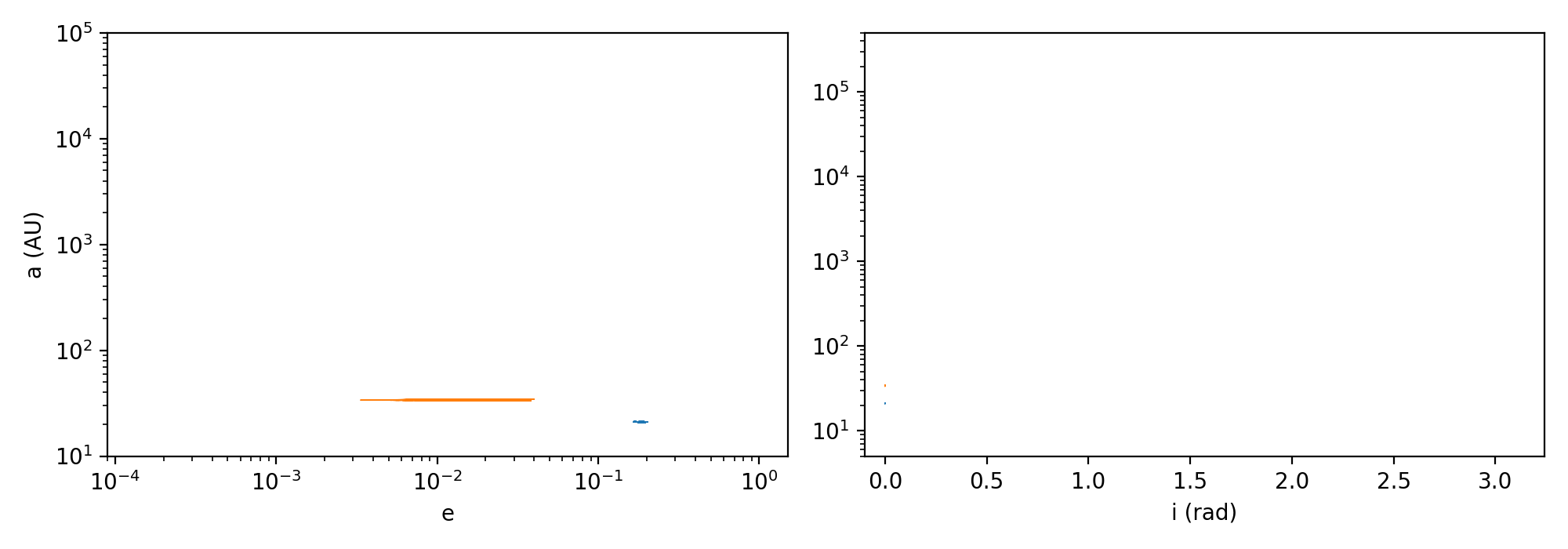}
        \caption{An isolated resonant \pds system. Both planets remain planar and oscillate around their initial semi-major axis and eccentricity.}
        \label{fig:parameters_pds_p_sys_isolated_0}
    \end{subfigure}
    \newline
    \begin{subfigure}[t]{\textwidth}
        \centering
        \includegraphics[keepaspectratio,width=\linewidth]{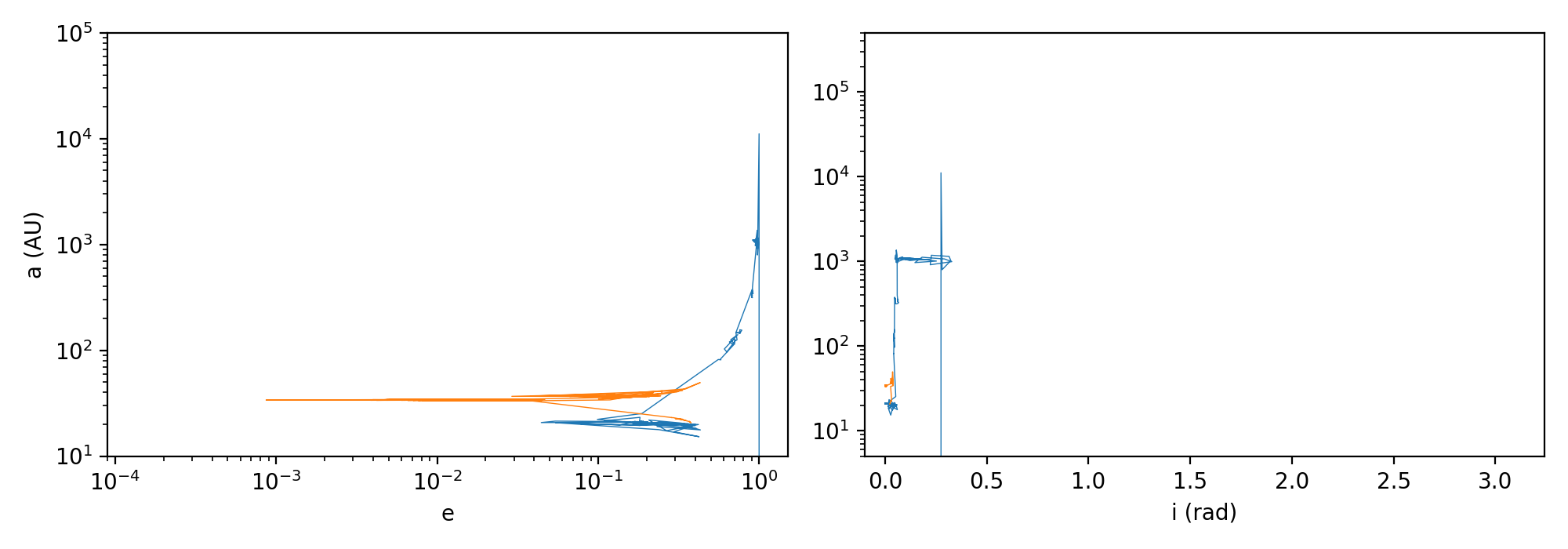}
        \caption{A perturbed resonant \pds system. The planets lose their MMR due to a perturbation and no stable orbit is found before a planet is lost.}
        \label{fig:parameters_pds_p_sys_34}
    \end{subfigure}
    \newline
    \begin{subfigure}[t]{\textwidth}
        \centering
        \includegraphics[keepaspectratio,width=\linewidth]{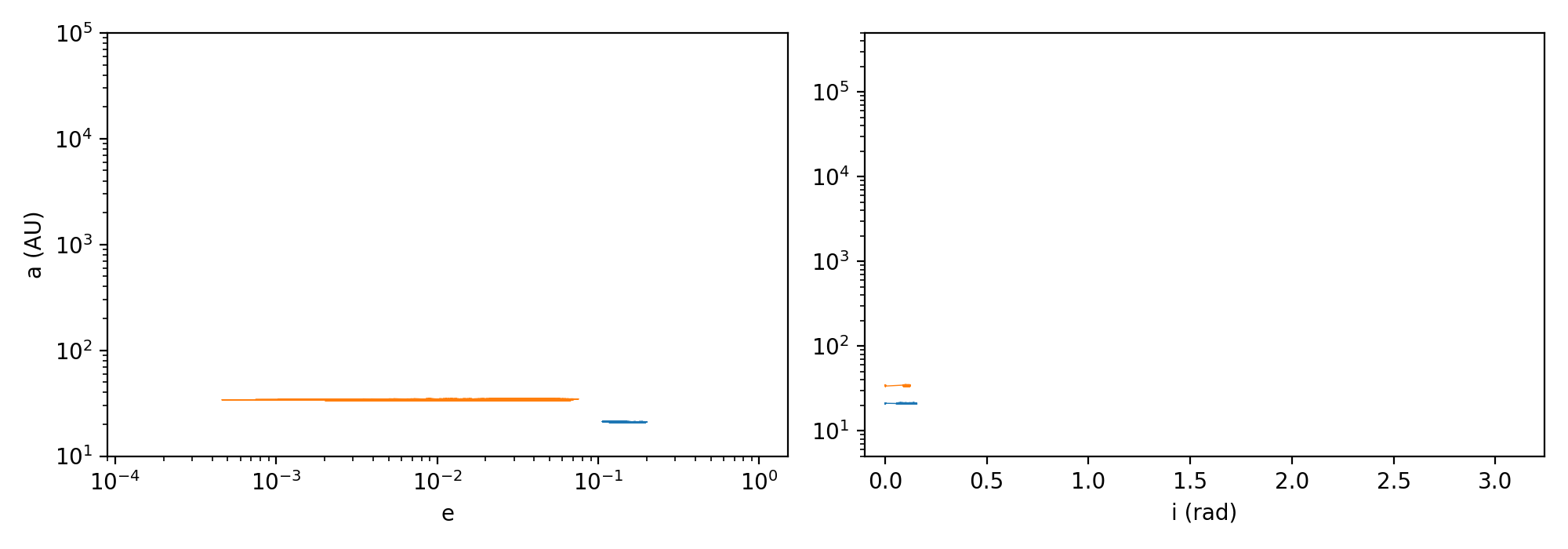}
        \caption{A perturbed resonant \pds system. The planets remain in approximately the same orbits after a perturbation, though tilted and oscillating at a greater magnitude.}
        \label{fig:parameters_pds_p_sys_15}
    \end{subfigure}
    \caption{The paths of semi-major axis versus eccentricity and semi-major axis versus inclination for a representative set of three different \pds systems.}
    \label{fig:parameters_pds}
\end{figure*}

%No evolution in MMR in isolation, negligible evolution in MMR in cluster with less noise than HR
Similar to \hr, the isolated resonant \pds system does not evolve other than some oscillation around its initial conditions (see figure \ref{fig:pds_arch_p_sys_isolated} and \ref{fig:parameters_pds_p_sys_isolated_0}). The data of the perturbed resonant systems in figure \ref{fig:pds_arch_p_sys} shows unstable perturbations of semi-major axes going up to 2000 AU, eccentricities up to 1, and inclinations seemingly short-lived above approximately 0.16 rad (for example, see figure \ref{fig:parameters_pds_p_sys_34}). The noise-like continuum of these perturbations that the perturbed resonant \hr systems displayed is far less present in the perturbed resonant \pds systems. The initial conditions remain the dominant mode, even after moderately strong perturbations (see figure \ref{fig:parameters_pds_p_sys_15}).

%Negligible evolution out of MMR in isolation with forbidden valley, negligible evolution out of MMR in cluster but now with inclinations and with overdensity from single system
The isolated non-resonant \pds systems in figure \ref{fig:pds_arch_p_sys_oor_isolated} show behaviour that is similar to that of the isolated non-resonant \hr systems: a dominant stable mode around the initial conditions and a noise-like continuum, but up to approximately 10000 AU rather than 3000 AU. Additionally, there is a forbidden valley of semi-major axes for \pds b around the initial orbit of \pds c. Qualitatively, the perturbed non-resonant systems do not differ much from the isolated non-resonant systems other than the existence of non-planar inclinations. The valley for \pds b around \pds c is less pronounced, the level of the noise-like continuum is comparable in width and height, and there are a few apparent stable modes belonging to a single system that maintained these modes for a prolonged amount of time.

%PDS negligibly affected by perturbations, so no divergence
The figures of \hr (figure \ref{fig:hr_arch}) and \pds (figure \ref{fig:pds_arch}) show that both \hr and \pds are affected by being non-resonant, but that only \hr is noticeably affected by the perturbations from neighbouring cluster stars as the variety of changes seen in figures \ref{fig:hr_arch_p_sys} and \ref{fig:hr_arch_p_sys_oor} is not present in figures \ref{fig:pds_arch_p_sys} and \ref{fig:pds_arch_p_sys_oor}. The non-resonant \pds systems produce similar distributions of semi-major axes up to 10000 AU since being out of resonance affects \pds far more than perturbations from neighbouring cluster stars. This explains why the divergence between isolated and perturbed non-resonant \hr systems seen in figure \ref{fig:survival_rate_hr} does not show up for isolated and perturbed non-resonant \pds systems in figure \ref{fig:survival_rate_pds}.

%Different cross-sections don't explain observations, so MMR chain increases susceptibility to external perturbations
One issue remains with this analysis: \hr and \pds have different interaction cross-sections. The observed difference between \hr and \pds can potentially be explained by wider planetary systems being affected more by the perturbations of neighbouring stars than less wide planetary systems. We can see in figures \ref{fig:hr_arch_p_sys} and \ref{fig:pds_arch_p_sys} that the noise-like continuum of the perturbed resonant \hr systems is roughly a factor 100 higher than that of the perturbed resonant \pds systems. The initial semi-major axis of \hr's outermost planet is about a factor 2 larger than that of \pds's outermost planet. This factor 2 in size equates to a factor 4 in cross-section, which means that \hr should have a factor 4 more interactions. This factor 4 cannot explain the factor 100 we observe in our data, which means that the difference in interaction cross-section between \hr and \pds cannot explain the difference in behaviour between \hr and \pds. We can only conclude that the number of planets, and therefore the presence of an MMR chain, increases the susceptibility to external perturbations since the number of planets is the only remaining qualitative difference between \hr and \pds.

\end{appendix}

\end{document}